\documentclass[sigconf, nonacm]{acmart}

\usepackage{amsmath,amssymb,amsfonts}
\usepackage{algorithmic}
\usepackage{graphicx}
\usepackage{textcomp}
\usepackage{xcolor}
\usepackage{booktabs}
\usepackage{tikz}
\usepackage{pgfplots}
\usepackage{siunitx}
\usepackage{diagbox}
\usepackage[linesnumbered,ruled,vlined]{algorithm2e}
\usepackage[all,pdf]{xy}
\usepackage{makecell}
\usepackage{bm}
\usepackage{caption}
\usepackage{subcaption}
\usepackage{adjustbox}
\usepackage{multirow}
\usepackage{float}
\usepackage{array}
\usepackage{calc}
\usepackage{balance}
\usepackage{hyperref}

\newenvironment{icompact}{
  \begin{list}{$\bullet$}{
    \parsep 0.5pt plus 0.5pt
    \partopsep 0.5pt plus 0.5pt
    \topsep 0.5pt plus 1pt minus 0.5pt
    \itemsep 0.5pt plus 0.5pt
    \parskip 0pt plus 1pt
    \leftmargin 0.15in}
       }
  {\normalsize\end{list}}

\newcommand{\bdpara}[2][\bdparaskip]{\vskip#1\noindent\textbf{#2}\ }
\newcommand{\bdsubpara}[1]{\textit{#1}\ }
\newtheorem{definition}{Definition}
\newtheorem{theorem}{Theorem}

\pgfplotsset{compat=1.16}
\usepgfplotslibrary{groupplots}
\usetikzlibrary{plotmarks,patterns,chains,calc}
\usetikzlibrary{}
\pgfplotsset{
tick label style={font=\small  },
label style={font=\bfseries\small  },
legend style={font=\bfseries\small },
nodes near coords style={font=\bfseries\small  },
}
\newcommand\sys{GenAllReduce\xspace}
\newcommand\algo{GenTree\xspace}
\newcommand\model{GenModel\xspace}
\newcommand{\allr}{AllReduce\xspace}

\newcounter{nop}
\newcommand{\nop}{\#processors\if\thenop0\footnote{The number of processors.}\stepcounter{nop}\fi\xspace}
\newcommand{\abc}{$(\alpha,\beta,\gamma)$\xspace}
\newcommand{\bwop}{\emph{bandwidth-optimal}\xspace}
\newcommand{\ltop}{\emph{latency-optimal}\xspace}
\newcommand{\op}[1]{\emph{#1}\xspace}
\newcommand{\cps}{{Co-located PS}\xspace}
\newcommand{\hcps}{{Hierarchical \cps}\xspace}
\newcommand{\ring}{{Ring Allreduce}\xspace}
\newcommand{\RHD}{{Recursive Halving and Doubling}\xspace}
\newcommand{\rb}{{Reduce-Broadcast}\xspace}
\newcommand{\sr}{\op{ReduceScatter}}
\newcommand{\ag}{\op{AllGather}}
\newcommand{\githuburl}{\url{https://anonymous.4open.science/r/AllreduceBenchmark-StreamEmulator-CCF4}}
\newcommand{\IE}{{i.e.}}
\newcommand{\EG}{{e.g.}}
\newcommand{\ETC}{{etc.}}
\newcommand{\naive}{na\"{\i}ve}

\newcommand{\del}[1]{\relax}
\newcommand{\lcomment}[1]{}
\newcommand{\optima}{}
\RequirePackage{graphicx, xcolor, amssymb}

\def\BibTeX{{\rm B\kern-.05em{\sc i\kern-.025em b}\kern-.08em
    T\kern-.1667em\lower.7ex\hbox{E}\kern-.125emX}}
\newcommand\vldbdoi{XX.XX/XXX.XX}

\begin{document}
\title{Revisiting the Time Cost Model of \allr}

\author{Dian Xiong}
\affiliation{%
  \institution{Tsinghua University}
  \state{Beijing, China}
}
\email{xd21@mails.tsinghua.edu.cn}

\author{Li Chen}
\affiliation{%
  \institution{Zhongguancun Laboratory}
  \state{Beijing, China}
}
\email{crischenli@gmail.com}

\author{Youhe Jiang}
\affiliation{%
  \institution{Tsinghua University}
  \state{Beijing, China}
}
\email{youhejiang@gmail.com}

\author{Dan Li}
\affiliation{%
  \institution{Tsinghua University}
  \state{Beijing, China}
}
\email{tolidan@tsinghua.edu.cn}

\author{Shuai Wang}
\affiliation{%
  \institution{Zhongguancun Laboratory}
  \state{Beijing, China}
}
\email{wangshuai@zgclab.edu.cn}

\author{Songtao Wang}
\affiliation{%
  \institution{Tsinghua University}
  \state{Beijing, China}
}
\email{kingplantwater@sina.com}

\begin{abstract}
AllReduce is an important and popular collective communication primitive, which has been widely used in areas such as distributed machine learning and high performance computing. To design, analyze, and choose from various algorithms and implementations of AllReduce, the time cost model plays a crucial role, and the predominant one is the $(\alpha,\beta,\gamma)$ model. In this paper, we revisit this model, and reveal that it cannot well characterize the time cost of AllReduce on modern clusters; thus must be updated. We perform extensive measurements to identify two additional terms contributing to the time cost: the incast term and the memory access term. We augment the $(\alpha,\beta,\gamma)$ model with these two terms, and present GenModel as a result. Using GenModel, we discover two new optimalities for AllReduce algorithms, and prove that they cannot be achieved simultaneously. Finally, striking the balance between the two new optimalities, we design GenTree, an AllReduce plan generation algorithm specialized for tree-like topologies. 
Experiments on a real testbed with 64 GPUs show that GenTree can achieve 1.22$\times$ to 1.65$\times$ speed-up against NCCL.
Large-scale simulations also confirm that GenTree can improve the state-of-the-art AllReduce algorithm by a factor of $1.2$ to $7.4$ in scenarios where the two new terms dominate.
\end{abstract}

\maketitle



\section{Introduction} \label{sec:introduction}

\allr is one of the well-known collective communication primitives in parallel computing. 
Conceptually, it can be considered as a concatenation of a \emph{reduce} operation, which collects data or partial results from different processors and combines them into a global result, and a subsequent \emph{broadcast} operation, which distributes the result to all processors.
\allr is also the most popular primitive~\cite{chunduri2018char-ar_cost_in_mpi, rabenseifner1999automatic}, and has seen wide usage in distributed machine learning (DML) and high performance computing (HPC). Therefore, improving the efficiency of \allr has gained continuous interest from both industry and academia~\cite{ringbaidu,geng2018hips,jiang2020unified,thakur2005optimization,2dhra}, and it has many different algorithms and implementations, such as Ring \allr~\cite{ringbaidu}, Parameter Server (PS)~\cite{geng2018hips,jiang2020unified} and Recursive Halving and Doubling (RHD)~\cite{thakur2005optimization}, \ETC 

To design, analyze, and choose from diverse algorithms and implementations, the time cost model of \allr plays a crucial role. 
The mainstream model for \allr is the \abc model~\cite{1994The}, and in this model, the \allr process is broken down into three parts: start-up, communication, and computation. The start-up cost is fixed, and it represents the latency of the communication, including the overheads of initiating a transfer, link delay, \ETC. The communication cost is related to the link bandwidth, while the computation cost is related to the computing power. These three costs are denoted by $\alpha$, $\beta$ and $\gamma$, respectively, hence the name~\cite{thakur2005optimization, 1994The, costmodel-1, costmodel-2, costmodel-3}. 

Previous efforts in optimizing \allr use this model in various ways: 
(1) specifying different optimality, such as latency optimal (\IE, the lowest start-up cost) and bandwidth optimal (\IE, the lowest communication cost)~\cite{patarasuk2009bandwidth};
(2) some work proposes new algorithms and proves their benefits with respect to the \abc model~\cite{geng2018hips,thakur2005optimization}, and some other work uses the model to generate algorithms automatically~\cite{SCCL,taccl}; 
(3) some communication libraries leverage the \abc model to choose which algorithms to use under a given situation~\cite{mpi40,mpich}. \par 

\begin{figure*}[t!]
    \centering
    \includegraphics[width=0.9\textwidth]{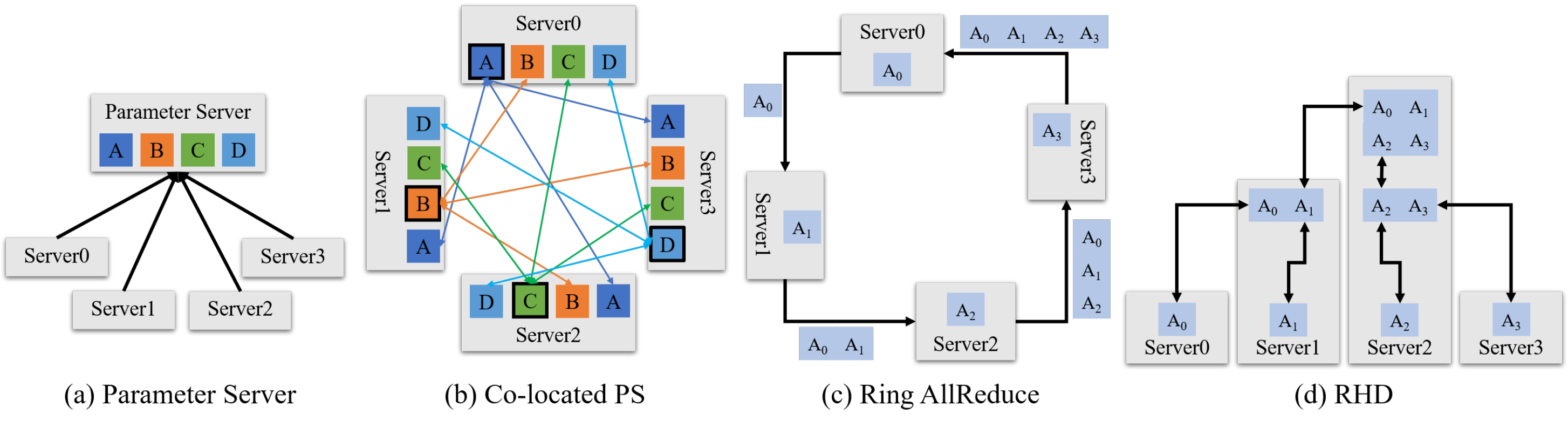}
    \caption{\allr plan types}
    \vspace{-0.2in}
    \label{fig:algs}
\end{figure*}

Despite the wide adoption of \abc model, we find that it cannot well characterize the time cost of \allr on modern clusters, thus must be updated. We argue that at least two more factors must be considered:
\begin{icompact}
    \item[1.] \textbf{Incast}: The incast problem is caused by bandwidth competition when machines perform many-to-one communication, \IE, the available link capacity is smaller than the aggregated bandwidth assumed by the \abc model. With the rapid increase of cluster size and the number of communication participants, the extra overhead introduced by incast is not negligible. Our experiments show that this problem exists even in Remote Direct Memory Access (RDMA) networks, and its severity is closely associated with the number of communicators (Section~\ref{ssec:designincast}). 
    \item[2.] \textbf{Memory access}: With the growth of per-host bandwidth, the gap between network bandwidth and memory bandwidth narrows~\cite{Jain2016recent}. Therefore, the memory access cost becomes non-negligible. An accurate time cost model must consider the memory access time before and after the computation (\emph{not} the memory copy between NIC and memory). Our analysis finds that this overhead is closely associated with the computation pattern of \allr algorithms (Section~\ref{ssec:memio}). 
\end{icompact} \par

Therefore, we augment the \abc model with two new terms: the incast term ($\varepsilon$) and the memory access term ($\delta$). The new model, dubbed \model, reflects the \textbf{gen}uine time costs of \allr on modern clusters. With \model, we can accurately predict the time cost of one \allr operation on a given network with detailed information of each factor's contribution. Compared to the \abc model, \model provides a better characterization of the performance of \allr. In our test scenarios, the maximum error of \model is \SI{2.6}{\percent}, while the \abc model is \SI{19.8}{\percent}. 

\model is a new perspective from which we can analyze \allr and design new algorithms. \model enables us to define two more optimalities: incast optimal (\IE, the lowest incast overhead) and memory access optimal (\IE, the lowest memory access cost). Our further analysis shows that these two optimalities \emph{cannot} be achieved simultaneously, and there must be a trade-off. 
We show that balancing these two optimalities can lead to new \allr algorithms, which can achieve superior performance in appropriate scenarios. 
Since the problem of generating the optimal \allr plan on arbitrary topology is proven to be NP-hard (Section~\ref{ssec:bg_topo}), in this paper, we focus on the widely used tree topology in DML and HPC deployments, and propose a heuristic plan generation algorithm, \algo, which strikes a balance between the two new optimalities. We show that \algo improves the state-of-the-art \allr algorithms on a real, small-scale testbed (Section~\ref{ssec:testbedexpr}), as well as in large-scale simulations (Section~\ref{ssec:evalsim}).


In summary, we make the following contributions:
\begin{icompact}
    \item We propose \model, an accurate time cost model for \allr. We show that \model is helpful in analyzing and designing new \allr algorithms. Tests show that \model can correctly predict the best algorithm while the \abc model cannot. In our test scenarios, \model's maximum error is \SI{2.6}{\percent}, whereas the \abc model's is \SI{19.8}{\percent}. We release an open-source benchmarking toolkit to help users fit \model to new clusters.
    \item We prove that generating the optimal \allr plan on an arbitrary topology is NP-hard. To demonstrate the usefulness of \model, we propose \algo, a heuristic scheme that generates \allr algorithms for the most widely used type of topology for DML and HPC.
    \item We implement \algo on both CPU and GPU testbeds to verify its performance benefits. Experiments show that \algo can generate \allr plans which achieve up to $2.4\times$ speedup compared to the state-of-the-art algorithms. 
    \item We developed a \allr simulator to test \model and \algo on a larger scale. Experiments show that \algo can produce \allr plan that can achieve $7.4\times$ speedup at max over the state-of-the-art algorithms. We also release the source code of the simulations to assist reproduction of our results.
\end{icompact} \par 
The rest of this paper is organized as follows. Section~\ref{sec:background} introduces the background and the motivation. Section~\ref{sec:genmodel} presents the details of \model. Section~\ref{sec:gentree} elaborates the algorithmic design of \algo based on \model. Section~\ref{sec:impl} describes the implementation and evaluation of \model and \algo in our testbed. Section~\ref{sec:related_work} outlines related works. Finally, Section~\ref{sec:conclusion} concludes the whole paper. \par 
This work does not raise any ethical issues.

\section{Background} \label{sec:background}
In this section, we first overview important \allr algorithms, and then introduce the \abc model to analyze them. Finally, we discuss the deficiencies of the \abc model, and provide motivations for a new one.

\subsection{Types of \allr Plan}\label{ssec:bg_allr}

\allr is the most popular collective communication primitive~\cite{chunduri2018char-ar_cost_in_mpi, rabenseifner1999automatic}. It reduces and synchronizes data among multiple processors. 
Recently, a notable application of \allr is DML~\cite{ringbaidu,jiang2020unified,jeaugey2017nccl,geng2018hips}, and researchers have demonstrated that the time cost of \allr operations accounts for nearly half of the total time cost in DML~\cite{luo2018parameter}. \par

We define an \allr \emph{plan} as an ordering of the data movement and reducing steps to complete an \allr primitive.
We overview four typical \allr plan types below. Performance is discussed in single-switch single-layer full-duplex networks where several servers are directly connected to one switch (for simplicity, we call this single-switch network later). We use $N$ for the number of processors and $S$ for the amount of data to \allr. \par

\bdpara{Parameter-Server-based Reduce-Broadcast.} A \naive{} way of \allr is \op{reduce} and then \op{broadcast}. \op{Reduce} means that processors send all their data to one server (which is called parameter server (PS)), and the PS aggregates data into one block. \op{Broadcast} means that the PS broadcasts the reduced data back to all processors. This type of plan is usually inefficient because the bandwidth and the computing power of the PS become the bottleneck and the resources of other processors are wasted. 

Thus, most practical \allr plans adopt the \sr-\ag strategy. Processors first partition data into $N$ blocks and each processor gathers and reduces one (\sr). Then each processor sends the block that it reduced to others (\ag) (see \cite{cho2019blueconnect} for more details). Most high-performance \allr plans are designed this way, as described below.


\bdpara{Co-located PS.} PSes and processors are co-located, as shown in Figure \ref{fig:algs}b. The whole data is partitioned into $N$ equal-sized blocks. Each processor works as a PS collecting one block. The communication pattern of \cps is balanced full-mesh since each processor sends/receives $2S(N-1)/N$ data to/from the other $(N-1)$ processors in total. Therefore, it can possibly leverage all the bandwidth resources in single-switch networks. However, many-to-one communications may lead to network congestion. Furthermore, in large networks, it will generate multiple flows and cross traffic thus possibly leading to PFC deadlock and spreading congestion problems~\cite{geng2018hips}. \par

\bdpara{Ring \allr.} \ring is widely used by DML frameworks~\cite{patarasuk2009bandwidth,ringbaidu,sergeev2018horovod}. Processors are arranged in a ring and only talk to their two neighbors, as shown in Figure \ref{fig:algs}c. The whole data are partitioned into $N$ blocks and \ring is finished in $2(N-1)$ steps. In step $j$\footnote{All numbers count from $0$}, processor $i$ will receive block-$((i - j)\mathbin{\%}N)$ from the left neighbor and send block-$((i - j+1)\mathbin{\%}N)$ to the right neighbor. \ring has a simple communication pattern, minimizes the inter-rack traffic and does not cause congestion. However, \ring has long dependency chains, leading to high latency, especially for large clusters. \par 

\bdpara{\RHD (RHD).} RHD constructs binary trees of processors, and processors communicate pairwise, as shown in Figure \ref{fig:algs}d. In step $0$ processors send/receive half of their data and in step $j$ send/receive $2^{-(j+1)}$ of their data. After $\log N$ steps, each processor has $S/N$ data already reduced. Processors then communicate opposite the previous $\log N$ steps. Two extra steps are needed if $N$ is non-power-of-two; therefore RHD consists of $2\lceil\log N\rceil$ steps. RHD is widely used in collective communication libraries such as MPI \cite{thakur2005optimization}. \par 

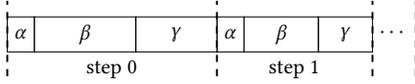
\begin{figure}[tb]
    \centering\optima
    \scalebox{0.9}{
    \begin{tikzpicture}[start chain,node distance=-0.15mm]
        \tikzstyle{box}=[draw,minimum height=0.52cm]
        \tikzstyle{sep}=[dashed, thick]
        \node [on chain, box, minimum width=0.4cm] (p1) {$\alpha$};
        \node [on chain, box, minimum width=1.5cm] {$\beta$};
        \node [on chain, box, minimum width=1.2cm] {$\gamma$};
        \node [on chain, box, minimum width=0.4cm] (p2) {$\alpha$};
        \node [on chain, box, minimum width=1.1cm] {$\beta$};
        \node [on chain, box, minimum width=0.8cm] {$\gamma$};
        \node [on chain, box, draw=none] (p3) {$\cdots$};
        \draw [sep] ($(p1.south west) + (0.007,-0.35)$)--($(p1.north west) + (0.007,0.23)$);
        \draw [sep] ($(p2.south west) + (0.007,-0.35)$)--($(p2.north west) + (0.007,0.23)$);
        \draw [sep] ($(p3.south west) + (0.007,-0.35)$)--($(p3.north west) + (0.007,0.23)$);
        \draw [sep] ($(p3.south east) + (0.007,-0.35)$)--($(p3.north east) + (0.007,0.23)$);
        \node at ($(p1.south west)!0.5!(p2.south west) - (0,0)$) [anchor=north] {step 0};
        \node at ($(p2.south west)!0.5!(p3.south west) - (0,0)$) [anchor=north] {step 1};
    \end{tikzpicture}
    }
    \caption{\allr in the view of the \abc model: in each step, first launching the transmission, then transmitting the data, finally aggregating the received data.}
    \label{fig:abc_sketch}
\end{figure}

\subsection{The \abc Model}\label{ssec:bg_abc}

The \abc model is the predominant model used in the analysis of collective algorithms \cite{thakur2005optimization, 1994The, costmodel-1, costmodel-2, costmodel-3}. It is formulated as
\begin{equation}\label{eq:abc}
    A\alpha + B\beta + C\gamma
\end{equation}
where $A$ is the communication rounds; $\alpha$ is a fixed cost that represents the latency of communication, including the overheads of initiating a transfer, link delay, \ETC; $B$ is the amount of data  transferred through a physical link; $\beta$ is the inverse bandwidth of the link and represents per-unit transmission costs (unit: byte, bit, or 4-byte float); $C$ is the number of aggregating operations (\EG, \texttt{sum} or \texttt{max}); $\gamma$ is the inverse CPU computation throughput and represents per-operation computation costs. In \allr, these three parts take place successively, as Figure \ref{fig:abc_sketch} shows. \par 

We can analyze \allr algorithms with this model. For example, in a single-switch network, the cost expressions for the four typical \allr plan types are listed in Table~\ref{tab:abcmodel_on_exist}.
\begin{table}[t]
    \small
    \centering
    \renewcommand{\arraystretch}{1.5}
    \caption{\abc model for some \allr plan types in single-switch networks. $\chi(x) = 0$ if $x$ is power-of-two else $\chi(x) = 1$.}
    \begin{tabular}{c|c}
        \hline
         \textbf{Type of Plan} & \textbf{\abc model expression} \\ \hline
         \rb & $2\alpha + 2(N-1)S\beta + 2(N-1)S\gamma$ \\ \hline
         \cps & $2\alpha+\frac{2(N-1)S}{N}\beta+\frac{(N-1)S}{N}\gamma$ \\ \hline
         \ring & $2(N-1)\alpha+\frac{2(N-1)S}{N}\beta+\frac{(N-1)S}{N}\gamma$ \\ \hline
         RHD  &  \makecell*[c]{$2\lceil\log N\rceil\alpha+\frac{2(N-1)S}{N}\beta+\frac{(N-1)S}{N}\gamma$\\[1.5pt] $+\chi(N)(2S\beta+S\gamma)$}\\ \hline
    \end{tabular}
    \label{tab:abcmodel_on_exist}
\end{table}
We can then infer several properties from the above analysis. First, \rb is much slower than the other three algorithms, as it wastes bandwidth and computing resources. Second, \cps and \rb have the lowest latency term, as they only have two steps. This property is called \ltop. Third, \cps and \ring have the lowest bandwidth term. This property is called \bwop. If $N$ is power-of-two, this property holds for RHD. Prior work~\cite{patarasuk2009bandwidth} has proved that, in \allr, the lowest traffic each processor sends to or receives from the network is 
\begin{equation}\label{eq:bw_op}
    2\frac{(N-1)S}{N}
\end{equation}
Therefore, an algorithm is \bwop if and only if the traffic to/from each processor is equal to the value.


\subsection{Motivation}\label{ssec:bg_motivation}
The \abc model can not accurately characterize the real overhead in modern clusters. We find that at least two more factors must be considered: incast and memory access, which gradually become non-negligible as networks grow larger and faster. \par 

\bdpara{Factor 1. Incast.}\label{ssec:bg_incast}
The incast problem is defined as the phenomenon that the actual bandwidth can not reach the theoretical link bandwidth when multiple flows congest the same link. Incast is well-known because it is severe in TCP~\cite{chen2009understanding_tcpincast, chen2015comprehensive_tcpincast}. Modern high-performance transport layer protocols such as RDMA also suffer from this problem~\cite{mittal2018revisiting-rdma, li2019hpcc}. \par

We show that the \abc model is inaccurate in incast scenarios, particularly for RDMA over Converged Ethernet (RoCE) networks. RoCE is the most commonly deployed RDMA technology~\cite{mittal2018revisiting-rdma}. In RoCE networks, the incast problem is closely related to the Priority Flow Control (PFC) mechanism. As loss recovery is too resource-intensive to handle in RoCE network interface cards (NICs), users usually enable PFC to achieve lossless delivery~\cite{mittal2018revisiting-rdma, li2019hpcc}. When the queue exceeds a certain threshold, the receiver will send pause frames to the upstream node, and the latter will pause the traffic to prevent buffer overflow~\cite{zhu2016ecn}. Therefore, PFC may lead to a bandwidth loss because all upstream links are blocked. Experiments in Section \ref{ssec:designincast} show that the growth trend of the pause frames is similar to that of the extra communication overhead. \lcomment{The incast overhead takes about \SI{8.2}{\percent} in the total \allr time of 12 processors.}In collective communications, with today's ever-expanding scale of parallel computing, the incast problem  becomes severe gradually and leads to unacceptable additional overhead. \par 

\bdpara{Factor 2. Memory Access.}\label{ssec:bg_memio}
Two possible processes in \allr involve memory access: communication and computation. During communication, memory copy occurs between the system kernel and the application, which RDMA can eliminate. During computation, the processor (CPU or GPU) needs to read from and write to memory. We note that the \abc model has an inadequate characterization of the memory access in computation. \par

To meet the rapid surge of network traffic, the bandwidth capacity of data center networks continues to increase. As the NIC bandwidth approaches the memory bandwidth in today's high-performance clusters \cite{Jain2016recent,nva100}, the memory access cost gradually becomes a non-negligible part of the overall \allr cost in high-speed networks. Later analysis (Section \ref{ssec:memio}) finds that the difference in the memory access overhead between algorithms can reach \SI{200}{\percent}.  \par

\bdpara{Need for a New Model.}
The above two factors render the \abc model inaccurate for modern clusters, and we expect the discrepancy to grow with the faster link speed as well as the larger size of networks. This inaccuracy prevents the \abc model from predicting the best \allr algorithm, as later we discuss in Section \ref{ssec:evalcm}. Therefore, we need to design a new cost model which takes them into account, helping us better understand the collective communication system. 
Using the new model, we can design performant \allr algorithms that can balance the different optimalities derived from the new model. Since generating the fastest \allr algorithm on any topology is NP-hard and therefore existing state-of-the-art solutions can not well handle large clusters, in Section~\ref{sec:gentree}, we design an algorithm that generates highly efficient \allr plans on the widely used tree-like physical topology of any size.

\section{\model: An Up-to-date \allr Time Cost Model}\label{sec:genmodel}
In this section, we describe the formulation and evaluation of \model. Compared to the \abc model, \model has two additional terms: the incast term and the memory access term. We first discuss the derivation of these two terms. Then we present the complete \model{\@}. Finally, we perform evaluations to demonstrate its accuracy and generality.

\bdpara{Experimental Settings.} The testbed has 15 servers connected to one single switch. Each server has dual 16-core \SI{2.4}{\GHz} Intel Xeon E5 Processors, \SI{128}{\gibi\byte} \SI{2400}{\MHz} DDR4 RAM, Mellanox RoCEv2 NIC and one NVIDIA K40c GPU. RDMA and PFC are enabled. The NICs and the switch are set to the speed of \SI{10}{Gbps} by default. We use \verb|float| as the data type. 1 \verb|float| occupies 4 bytes, and \SI{400}{\mega\byte} means 100 million \verb|float|s for example. When using MPI, timestamps are obtained by \texttt{MPI\_Wtime} and experiments are repeated 100 times and the mean values are taken. \par

\subsection{The Memory Access Term ($\delta$)} \label{ssec:memio}
We focus on memory access during computation. Different \allr algorithms may still generate different memory access overheads for the same input and output. Take \ring as an example. In each step, one processor receives one piece of data and computes one-by-one, expressed as
\begin{equation}\label{eq:memio1}
a_0=a_0+a_1,\ a_0=a_0+a_2,\ \ldots,\ a_0=a_0+a_{N-1}
\end{equation}
where $a_i$ represents one piece of data on processor $i$ and there are total $N$ processors. Each sum operation involves two memory read operations and one memory write operation, so a total of $3(N-1)$ memory read/write steps are required. As a comparison, in PS, the root processor receives $(N-1)$ piece of data and computes only once, expressed as
\begin{equation}\label{eq:memio2}
a_0=a_0+a_1+...+a_{N-1}
\end{equation}
This involves $N$ memory read operations and $1$ memory write operation, and only a total of $(N+1)$ memory read/write steps are required. Thus, the number of memory r/w steps relates directly to the computation fan-in degree. The degree of the Ring-like computation pattern is $2$, and the PS-like is $N$. \par 

\begin{figure}[t]
    \centering
    \footnotesize
    \begin{tikzpicture}
        \begin{axis}[
            width=0.9\linewidth,
            height=0.6\linewidth,
            axis y line*=right,
            axis x line=none,
            ylabel={extra comm overhead (\si{\s})},
            ymin=-0.00875, ymax=0.14,
            xmin=5, xmax=16,
            every axis y label/.append style={ACMDarkBlue},
            y tick label style={
                /pgf/number format/.cd,
                    fixed,
                    fixed zerofill,
                    precision=2,
                /tikz/.cd
            },
            ylabel shift=-3pt,
            ytick distance=0.04,
            ]
            \addplot+[color=ACMDarkBlue, mark size=1.5pt, mark options={ACMDarkBlue}, thick, mark=square*]  coordinates {
                (6,0) (7,0) (8,0) (9,0) (10,0.005) (11,0.023) (12,0.052) (13,0.08) (14,0.1) (15, 0.13)
            };
        \end{axis}
        \begin{axis}[
            width=0.9\linewidth,
            height=0.6\linewidth,
            xlabel={Full-mesh \#nodes},
            ylabel={\#pause frames ($10^4$)},
            axis y line*=left,
            ymin=-1, ymax=16,
            xmin=5, xmax=16,
            y label style={ACMRed},
            ylabel shift=-3pt,
            ytick distance=4,
            minor y tick num=1,
            legend entries={\#pause frames, extra comm overhead},
            legend pos={north west},
            ]
            \addplot+[color=ACMRed, mark size=1.5pt, mark options={ACMRed}, thick] coordinates {
                (6,0) (7,0) (8,0) (9,0) (10,1.8) (11,4) (12,6.2) (13,9.05) (14,11.9) (15, 15)
            };
            \addplot+[color=ACMDarkBlue, mark size=1.5pt, mark options={ACMDarkBlue}, thick, mark=square*]  coordinates {(0,0)};
            \addplot+[color=ACMDarkBlue, no marks, thick, domain=6:15.2]  {0};
            \node[anchor=west] at (15.1,0.1) {\textcolor{ACMDarkBlue}{$\beta$}};
            \node[rotate=90, anchor=west] at (14.5,0) {\textcolor{ACMDarkBlue}{\scriptsize          Discrepancy of}};
            \node[rotate=90, anchor=west] at (15.0,0) {\textcolor{ACMDarkBlue}{\scriptsize          \abc Model}};
            \draw[<->, >=stealth, color=ACMDarkBlue, thick] (14, 0.2) -- (14, 11);
        \end{axis}
    \end{tikzpicture}
    \caption{PFC pause frames and extra communication overhead of $x$-to-$1$ communications with $x$ ranging from $6$ to $15$.}
    \label{fig:incast}
\end{figure}
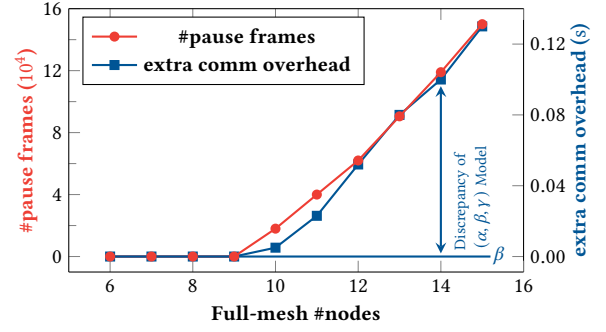
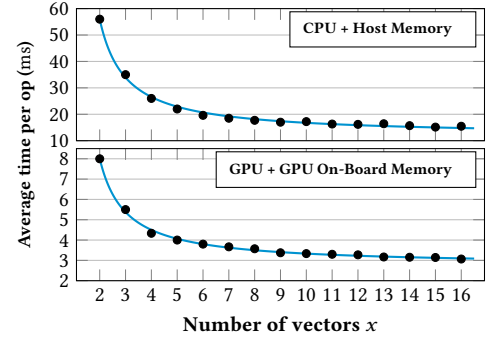
\begin{figure}[t]
    \small
    \def\mystrut{\scriptsize\vphantom{hg}}
    
    \pgfplotsset{
        legend image with text/.style={
            legend image code/.code={%
                \node[anchor=center] at (0.3cm,0cm) {\bfseries #1};
            }
        },
    }
    \centering
    \begin{tikzpicture}
        \begin{groupplot}[
            group style={
                group size=1 by 2,
                vertical sep=3pt,
                x descriptions at=edge bottom,
                ylabels at=edge left,
            },
            xlabel={Number of vectors $x$},
            xtick distance=1,
            scale only axis,
            ymajorgrids,
            width=0.64\linewidth,
            height=50pt,
            ylabel shift=-3pt,
            xtick align=inside,
            legend style={
                font=\mystrut,
                legend cell align=center,
            },
        ]
        \nextgroupplot[
            ytick distance=10,
            xmin=1.1, xmax=16.9,
            ymin=10, ymax=60,
        ]
        \addlegendimage{legend image with text={CPU + Host Memory}}
        \addlegendentry{}
        \addplot+[thick, domain=2:16.5, no marks, samples=1000, ACMBlue] {44.07/(x-1) + 11.87};
        \addplot[only marks, mark size=1.5pt] table
        {
            2	56
            3	35
            4	26
            5	22
            6	19.6
            7	18.5
            8	17.71428571
            9	17
            10	17.22222222
            11	16.3
            12	16.18181818
            13	16.41666667
            14	15.69230769
            15	15.14285714
            16	15.46666667
        };
        \coordinate (top) at (rel axis cs:0,1);
        \nextgroupplot[
            ytick distance=1,
            xmin=1.1, xmax=16.9,
            ymin=2, ymax=8.5, 
        ]
        \addlegendimage{legend image with text={GPU + GPU On-Board Memory}}
        \addlegendentry{}
        \addplot+[thick, domain=2:16.5, no marks, samples=1000, ACMBlue] {5.257/(x-1) + 2.749};
        \addplot[only marks, mark size=1.5pt] table
        {
            2	8
            3	5.5
            4	4.333333333
            5	4
            6	3.8
            7	3.666666667
            8	3.571428571
            9	3.375
            10	3.333333333
            11	3.3
            12	3.272727273
            13	3.166666667
            14	3.153846154
            15	3.142857143
            16	3.066666667
        };
        \coordinate (bot) at (rel axis cs:1,0);
        \end{groupplot}
        \path (top-|current bounding box.west)--
        node[anchor=south,rotate=90] {\optima\bfseries\footnotesize Average time per op (\si{\ms})}
        (bot-|current bounding box.west);
    \end{tikzpicture}
    \caption{Average reduce overhead between every two vectors (${T_x}/(x-1)$) while processing $x$ 150M-float-vectors. The more vectors that are reduced at once, the faster each reduce operation will be. }
        \label{fig:calc}
\end{figure}

Following this rule, we define $\delta$ as the per-unit memory read/write time cost. The total memory access overhead is formulated as $D\delta$, where $D$ represents the amount of the memory operations. For example, \ring has $(N-1)$ computation steps, and each processor adds two blocks of $S/N$ data once. Its memory access cost is $3(N-1)\frac{S}{N}\delta$.
In contrast, \cps has only one computation step and each processor adds $N$ blocks of $S/N$ data. Its memory access cost is $(N+1)\frac{S}{N}\delta$.
Therefore, when $N$ is large, the difference of the memory access overhead between \cps and \ring can even reach \SI{200}{\percent}.

We take a micro-benchmark of memory access cost by adding $x=2,3,\dots,N$ vectors at once on one single machine of our testbed, and is done by C++ (CPU results) and CUDA (GPU results). Each vector consists of $S=\SI{150}{M}$ floats. This involves $(x+1)S$ memory operations and $(x-1)S$ \texttt{add} operations (\IE, the $\gamma$ term). Therefore, the time cost should be
\begin{equation}\label{eq:calc_bench}
\begin{split}
    T(x) &= (x+1)S\delta + (x-1)S\gamma \\
    \implies\ \frac{T(x)}{x-1} &= \frac{x+1}{x-1}C_1 + C_2
\end{split}    
\end{equation}
where $C_1$ (${}=S\delta$) and $C_2$ (${}=S\gamma$) are constants, and $\frac{T(x)}{x-1}$ is the average time cost of per-\text{add} operation. 
Figure \ref{fig:calc} shows the results, with black marks representing benchmark and blue lines being trend lines fitted according to Equation (\ref{eq:calc_bench}). This supports our analysis, and the memory cost can be saved by \SI{66.7}{\percent} at max when $x$ is large.

\subsection{The Incast Term ($\varepsilon$)} \label{ssec:designincast}
To understand incast, we perform $x$-to-$x$ communication tests with $x=2,3,\dots,N$ communicators (\IE, full-mesh, which is exactly what \cps does). Every communicator receives a fixed amount of data $S$. If there was no incast, the time cost should be
\begin{equation}
    T(x) = \alpha + S\beta
\end{equation}
which is a constant. We perform tests to confirm the validity of this formula using Open MPI with $S$ taking the value of \SI{20}{M}. Results reveal that this property holds when $2\leqslant x \leqslant 9$, while extra overhead emerges when $x$ is greater than $9$. Our further investigation reveals a potential relationship between the extra overhead and PFC pause frames, as shown in Figure~\ref{fig:incast}.

In summary, the incast problem relates directly to the fan-in degree ($=x$) of many-to-one communications, and we believe this is related to PFC. Below a certain threshold, which is denoted by $w_t$, no incast is observed; beyond the threshold, the extra overhead caused by incast grows linearly (we believe that linear approximation is sufficient), and the slope is denoted by $\varepsilon$. The incast overhead is formulated as 
\begin{equation}
    \max(w-w_t,0)B\varepsilon
\end{equation}
where $w$ is the communication fan-in degree and $B$ is the total amount of data received. \par

Taking PS-based \allr as an example. The root node receives data of size $S$ from each of the other nodes; the total communication cost is
\begin{equation}
    T = \alpha + (N-1)S\beta + \max(N-w_t, 0)(N-1)S\varepsilon
\end{equation}
Intuitively, $\varepsilon$ can be considered as a correction of the bandwidth coefficient $\beta$, as shown below.
\begin{gather}
    T = A\alpha + B\beta + \max(w-w_t, 0)B\varepsilon = A\alpha + B\beta'\\
    \beta' := \beta + \max(w-w_t,0)\varepsilon
\end{gather}

\subsection{\model and Its Implications} \label{ssec:fmltmodel}
Augmenting the \abc model with the above two terms, we obtain the \model, which can be formulated as:
\begin{equation}
    T = A\alpha + B\beta + C\gamma + D\delta + \max(w-w_t, 0)B\varepsilon.
\end{equation}

\begin{figure}[t]
    \centering
    \includegraphics[width=0.8\linewidth]{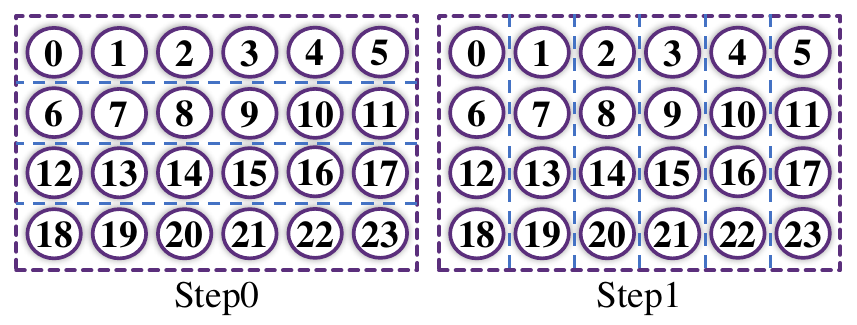}
    \caption{Example of $6\times4$ \hcps. In the first step, servers form 6-server groups, and do \sr inside the groups. In the second step, servers form 4-server groups, and do \sr on the results of the previous step inside the groups. These two groupings are orthogonal. }
    \label{fig:hcps24}
\end{figure}

In a single-switch network, the \model expressions for some typical \allr plan types are given in Table \ref{tab:genmodel_on_exist}. We can see how the two new terms differ among algorithms. Hierarchical Co-located PS (\hcps) is also included, with $m$ representing the number of steps and $f_i$ the fan-in degree of step-$i$. Steps grouping are orthogonal to each other. Figure~\ref{fig:hcps24} shows an example of $m=2$ and $f_0=6,f_1=4$. HCPS is quite useful in balancing the two new terms, which will be discussed later. In Section~\ref{sec:gentree}, we use the time cost models in Table \ref{tab:genmodel_on_exist} to select appropriate \allr plans.
\begin{table}[t]
    \optima
    \small
    \centering
    \renewcommand{\arraystretch}{1.5}
    \caption{\model for some \allr plan types in single-switch networks. Recall that $\chi(x) = 0$ if $x$ is power-of-two else $\chi(x) = 1$. For \hcps, $m$ is the number of steps and $f_i$ is the fan-in degree in step $i$.}
    \label{tab:genmodel_on_exist}
    \begin{adjustbox}{max width=\linewidth}
        \begin{tabular}{c|c}
        \hline
         \textbf{Type of Plan} & \textbf{\model expression} \\ \hline
         \makecell*[c]{\rb} & \makecell*[c]{$2\alpha + 2(N-1)S\beta + (N-1)S\gamma + (N+1)S\delta$ \\ + ${} 2(N-1)S\cdot\max(N-w_t,0)\varepsilon$} \\ \hline
         \ring & {$2(N-1)\alpha+\frac{2(N-1)S}{N}\beta+\frac{(N-1)S}{N}\gamma + \frac{3(N-1)S}{N}\delta$}\\ \hline
         RHD  & \makecell*[c]{$2\lceil\log N\rceil\alpha+\frac{2(N-1)S}{N}\beta+\frac{(N-1)S}{N}\gamma + \frac{3(N-1)S}{N}\delta$\\ ${}+\chi(N)(2S\beta+S\gamma+3S\delta$)}\\ \hline
         \cps & \makecell*[c]{$2\alpha+\frac{2(N-1)S}{N}\beta+\frac{(N-1)S}{N}\gamma + \frac{(N+1)S}{N}\delta$ \\ ${}+ \frac{2(N-1)S}{N}\max(N-w_t, 0)\varepsilon$}\\ \hline
         \makecell*[c]{Hierarchical\\\cps}& \makecell*[c]{$2m\alpha+\frac{2(N-1)S}{N}\beta+\frac{(N-1)S}{N}\gamma +\frac{2\sum_{i=1}^{m-1}(\prod_{j=1}^{i}f_{j})+N+1}{N}S\delta$ \\ $+\sum_{i=0}^{m-1}\left(\max(0, f_i-w_t)  \frac{(f_{i-1}-1)\prod_{j=i}^{m-1}f_j}{N}\right)S\varepsilon$}\\ \hline
    \end{tabular}
    \end{adjustbox}
\end{table}

\renewcommand{\arraystretch}{1}

As an example, we demonstrate how to use \model to analyze time cost of \allr plans with Hierarchical Co-located PS (\hcps). A $m\times n$ \hcps operates as follows. The total number of servers is $N=m\times n$. Firstly, servers form groups of size $m$, and do \sr within each group. Next, servers form groups of size $n$ with servers, and the new grouping is orthogonal to the previous grouping, \IE, each group does not contain servers from the same group in the prior step. With the new groups, servers perform \sr again. Finally, \ag is performed reversely to distribute the results. Figure~\ref{fig:hcps24} shows an example of $6\times 4$ \hcps. Formally, for \hcps the time cost is shown in Table \ref{tab:genmodel_on_exist}. \model can help us infer that 1) when using \hcps, the larger the prior steps' fan-in degrees, the less the memory access overhead; 2) the incast term is obtained by stacking each step's incast overhead, and if all $f_i$ are less than $w_t$, this term should be zero.

We proceed to prove two results immediately following \model. First, we define two new optimalities, and then reveal their respective lower bounds. Finally, we present an impossibility result that states the two optimalities cannot be achieved simultaneously.

\subsubsection{Incast Optimal}

\begin{definition}[$\varepsilon$-optimal]
Incast optimal means an \allr plan has the lowest incast overhead.
\end{definition}

Evidently, the lower bound for the $\varepsilon$ term is zero, as \ring generates no competing flows and thus avoids incast. 

\subsubsection{Memory Access Optimal}
\begin{definition}[$\delta$-optimal]
Memory access optimal means an \allr plan has the lowest memory cost.
\end{definition}

Next, we prove the lower bound for the $\delta$ term.

\begin{theorem} \label{theo:memio}
    The lower bound of memory access cost is 
    \begin{equation}\label{eq:memlowerbound}
        \frac{(N+1)S}{N}\delta
    \end{equation}
    One algorithm is memory access optimal if and only if its memory access cost is this value. 
\end{theorem}

\bdsubpara{Proof.} As there are a total of $N$ processors, each processor should collect and reduce one block of data ($S/N$) to maximize parallelism. Taking an arbitrary block of data, initially, all processors have this data block of their own; finally, only one processor has this data block that is already reduced globally, which is obtained by a sequence of computation operations $O_0, O_1, \dots, O_{h-1}$ ($h\geqslant1$). The fan-in degree of $O_i$ is denoted by $f_i$, \IE, $O_i$ reduces $f_i$ data blocks to one block. As there are $N$ data blocks initially and $1$ block finally, we can infer that
\begin{equation}\label{eq:mem1}
    N-1 = \sum_{i=0}^{h-1} (f_i-1) = -h + \sum_i f_i
\end{equation}
At the same time, $O_i$ reads $f_i$ data blocks from memory and then writes $1$ data block back to memory, so the total memory access overhead is 
\begin{equation}\label{eq:mem2}
    T = \sum_{i=0}^{h-1} (f_i + 1)\times\frac{S}{N}\delta = (h+\sum_i f_i)\times\frac{S}{N}\delta
\end{equation}
Substituting Equation (\ref{eq:mem1}) into Equation (\ref{eq:mem2}), we obtain
\begin{equation}\label{eq:mem3}
    T = (N-1+2h)\frac{S}{N}\delta
\end{equation}
Therefore, the more intermediate steps, the more the memory access overhead. Substitute $h=1$ to Equation (\ref{eq:mem3}), we obtain the result Equation (\ref{eq:memlowerbound}). \hfill $\square$ \par

\subsubsection{An Impossibility Result} \label{sssec:impossible}
\begin{theorem}\label{theo:impossible}
An \allr plan cannot be $\varepsilon$-optimal and $\delta$-optimal simultaneously for a network where the number of servers $N$ is greater than the incast threshold $w_t$.
\end{theorem}

\bdsubpara{Proof.} Each step of \allr must contain both communication and computation, as the output of the last step's computation is the input of the next step's communication. If an \allr plan is memory access optimal, \IE, only one step of computation happens on the server that has received $N-1$ data blocks from all other servers. This leads to the incast problem because $N>w_t$. On the other hand, If an \allr plan is incast optimal, \IE, $f_i\leqslant w_t< N$. From Equation (\ref{eq:mem1}), we know $h\geqslant2$, so it cannot be memory access optimal. \hfill $\square$ \par

Our experiments indicate that the incast threshold $w_t$ is less than 10 for our RoCE NICs, and we expect this number to also be small for other models of RoCE NICs. Thus, the number of servers usually exceeds $w_t$ for large-scale DML and HPC scenarios, and users of \allr must make a trade-off between the two optimalities.

From Theorem~\ref{theo:impossible}, we can also draw the insight that, to reduce the overall time cost, we can moderately increase the fan-in degree without incurring incast. Basic algorithms such as \ring, RHD, and \cps cannot benefit from this because their fan-in degrees are fixed at $2$, $2$, $N$, respectively. In Section~\ref{sec:gentree}, we describe how \algo can balance these two optimalities.

\subsection{Fitting \model to a New Cluster}\label{ssec:fitting}
Here we briefly describe how to measure the parameters in \model for a new cluster.
\model has six parameters to fit: $\alpha$, $\beta$, $\gamma$, $\delta$, $\varepsilon$, and $w_t$. We suggest running the \cps benchmark, which we provide as a part of the toolkit. These parameters can be fitted by feeding \cps benchmark results on $2,3,\dots,\mathit{max}$ communicators. However, according to Table \ref{tab:genmodel_on_exist}, the ratio of the $\beta$-term coefficient to the $\gamma$-term coefficient is always 2, which means we only need $(2\beta + \gamma)$. This is sufficient for the end-to-end time cost analysis of the \allr. If users must know the exact values of these two parameters, they can calculate $\beta$ from bandwidth and then calculate $\gamma$. 

\section{\algo: an \allr Plan Generation Algorithm}\label{sec:gentree}
In this section, we demonstrate the usefulness of \model in terms of \allr plan generation. We first examine the complexity of the problem of plan generation, and prove its NP-Hardness on arbitrary topology. 
Then we focus on tree-like topology, which is commonly used in practice, and present \algo, a \allr plan generation algorithm. 
We describe how \algo utilizes \model to enable a highly efficient \allr plan.

\subsection{NP-Hardness of \allr Plan Generation}\label{ssec:bg_topo}
The application scenario of \allr is usually much more complex than single-switch networks. The network topology may be a tree, a ring, or even several geographically distributed data centers. On these topologies, the \allr algorithm needs to be customized to achieve better performance~\cite{geng2018hips}. Unfortunately, obtaining an optimized \allr plan for arbitrary topology is not easy. \par

\begin{theorem}\label{theo:np}
    Generating an \allr plan with minimal makespan on arbitrary topology is NP-hard. 
\end{theorem}

\bdsubpara{Proof.} On a given physical topology $G=(V,E)$, let $N$ be the number of vertexes and $L$ the number of edges. Generating the optimal \op{AllGather} scheme on $G$ (as a part of \allr) can be translated into a job-shop scheduling problem as follows:
\begin{icompact}
    \item[1.] An edge between vertex $i$ and $j$ is transformed into machine(s). If it represents a half-duplex link, it is transformed into one machine $M_{ij}$. If it represents a full duplex link, it is transformed into two machines $M_{i\to j}$ and $M_{j\to i}$. 
    \item[2.] All vertex broadcast data in \op{AllGather}. Let $J_i$ ($i\in(0, N-1)$) represent the broadcast job sourced from vertex $i$. 
    \item[3.] $J_i$ consists of $N$ operations $O_{ij}$ ($j\in(0,N-1)$). $O_{ij}$ represents the operation that the data sourced from vertex $i$ is sent to vertex $j$.
    \item[4.] $O_{i0}, O_{i1}, \dots, O_{i(N-1)}$ are sequence-dependent, as data must reach one vertex's adjacency before reaching the vertex. 
    \item[5.] Operation $O_{ij}$ can be only conducted on machine $M_{*j}$, $M_{j*}$ or $M_{*\to j}$.
\end{icompact}
It is known that the job-shop problem with a sequence-dependent setup is NP-hard~\cite{sotskov1995np}. Therefore, finding the \allr plan with minimal makespan is also NP-hard. \hfill $\square$

\subsection{The Design of \algo} 
\label{ssec:hallr}
Given the hardness of the problem, we restrict the problem space to the widely used tree topology in DML and HPC deployments and design a heuristic algorithm to generate \allr plans with the help of \model. Although the plan generation problem is still complicated on tree topology, we leverage the fact that a tree shares similarity with the traffic pattern of the \op{Reduce} primitive: a tree has a root, and so does \op{Reduce}. We may build an efficient \sr plan under the guidance of \model; then \ag can be performed in the reverse order. By combining \sr and \ag, we obtain a complete \allr plan. Due to the symmetry between \sr and \ag, we only discuss the plan generation of \sr in the following, and we can reverse the \sr plan to obtain the \ag plan directly. \par

\begin{figure}[t]
    \centering
    \begin{subfigure}{0.4068\linewidth}
        \centering
        \includegraphics[width=0.8\linewidth]{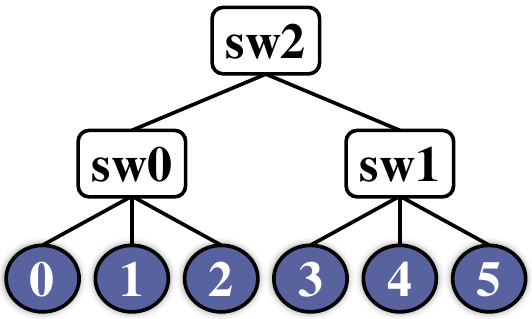}
        \caption{Symmetric tree}
        \label{fig:phytopo6}
    \end{subfigure}
    \begin{subfigure}{0.4762\linewidth}
        \centering
        \includegraphics[width=0.8\linewidth]{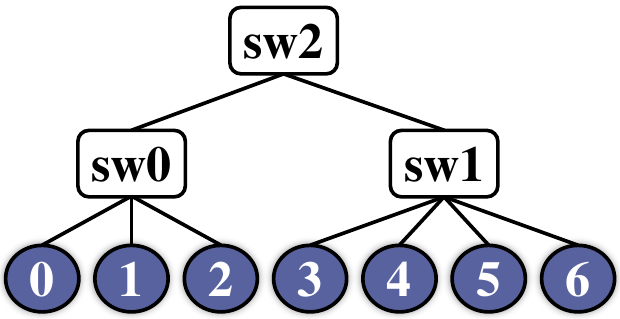}
        \caption{Asymmetric tree}
        \label{fig:phytopo7}
    \end{subfigure}
    \caption{Two examples of physical topology. ``sw'' is short for ``switch''.}
    \label{fig:phytopo}
\end{figure}

\bdpara{Understanding tree topology.} As shown in Figure~\ref{fig:phytopo}, each tree-based physical topology has a root node, and every non-root node has a link connecting to its parent. Each node can have an arbitrary number of children. The leaves of a tree are servers where data are stored and processed. Other non-leaf nodes are switches.
For FatTree~\cite{fattree} topology and Leaf-Spine topology~\cite{leaf-spine}, we choose a random top-level switch as the root and ignore the other top-level switches. Because we only care about the data movement between the servers, the choice of root in FatTree and Leaf-Spine topology does not affect the output of the \algo.

\bdpara{Algorithm design.} \algo is a recursive algorithm, which generates \sr plan for tree topology using \model. 
A \sr plan decides the data movement between the servers and the order of \op{Reduce} operations. A \algo-generated plan is a sequence of sub-plans, and each sub-plan describes the data movement and order of \op{Reduce} under a switch in the tree, which we call a switch-local sub-tree.

On a high level, for each switch-local sub-tree, \algo first generates a straightforward \emph{basic sub-plan}, and then uses \model to determine the optimal \emph{final sub-plan}.
Since each switch's \sr plan must depend on the initial data placement resulting from the switch's children's \sr sub-plans, \algo must work in a bottom-up, recursive manner. 
Finally, we collect all sub-plans to produce the \sr plan for the tree topology.  

We then elaborate on the detailed operations of \algo.

\bdpara{Basic sub-plan generation.} Consider a tree topology that consists of $N$ servers and several switches. Each server's data are split into $N$ blocks. \algo first generates a basic \sr sub-plan from the bottom layer of switches to the root switch of the topology. 
Consider a switch $A$ with $c$ children denoted by $C_i$ ($i\in\{0,1,\dots,(c-1)\}$).
The children can be switches or servers.
For a sub-tree with $C_i$ as the root node, there are total $n_i$ servers (leaves). 
Before \sr on $A$ can start, $C_0, C_1, \dots, C_{c-1}$ must finish their respective \sr operations. This means that, in the sub-tree with $C_i$ as the root node, each of its $n_i$ servers has finished the \op{Reduce} operation on $\lceil N/n_i\rceil$ blocks of data. If $C_i$ is a server (leaf), its \sr operation is considered done. 
After \sr of $A$ is done, in the sub-tree with $A$ as the root node, there are total $n=\sum_i n_i$ servers, and each of them has collected and reduced $\lceil N/n\rceil$ block(s) of data.
In this way, for each switch-local sub-tree, we know the initial data placement (before \sr) and the final data placement (after \sr), which enables us to produce a basic \sr sub-plan. In this basic sub-plan, for each data block, we directly move it from where it is stored (initial placement) to where it is reduced (final placement).

\begin{figure}[!tb]
    \centering
    \begin{subfigure}{\linewidth}
        \centering
        \includegraphics[width=0.74\linewidth*\real{1}, height=(\linewidth*\real{0.9065}*\real{1})]{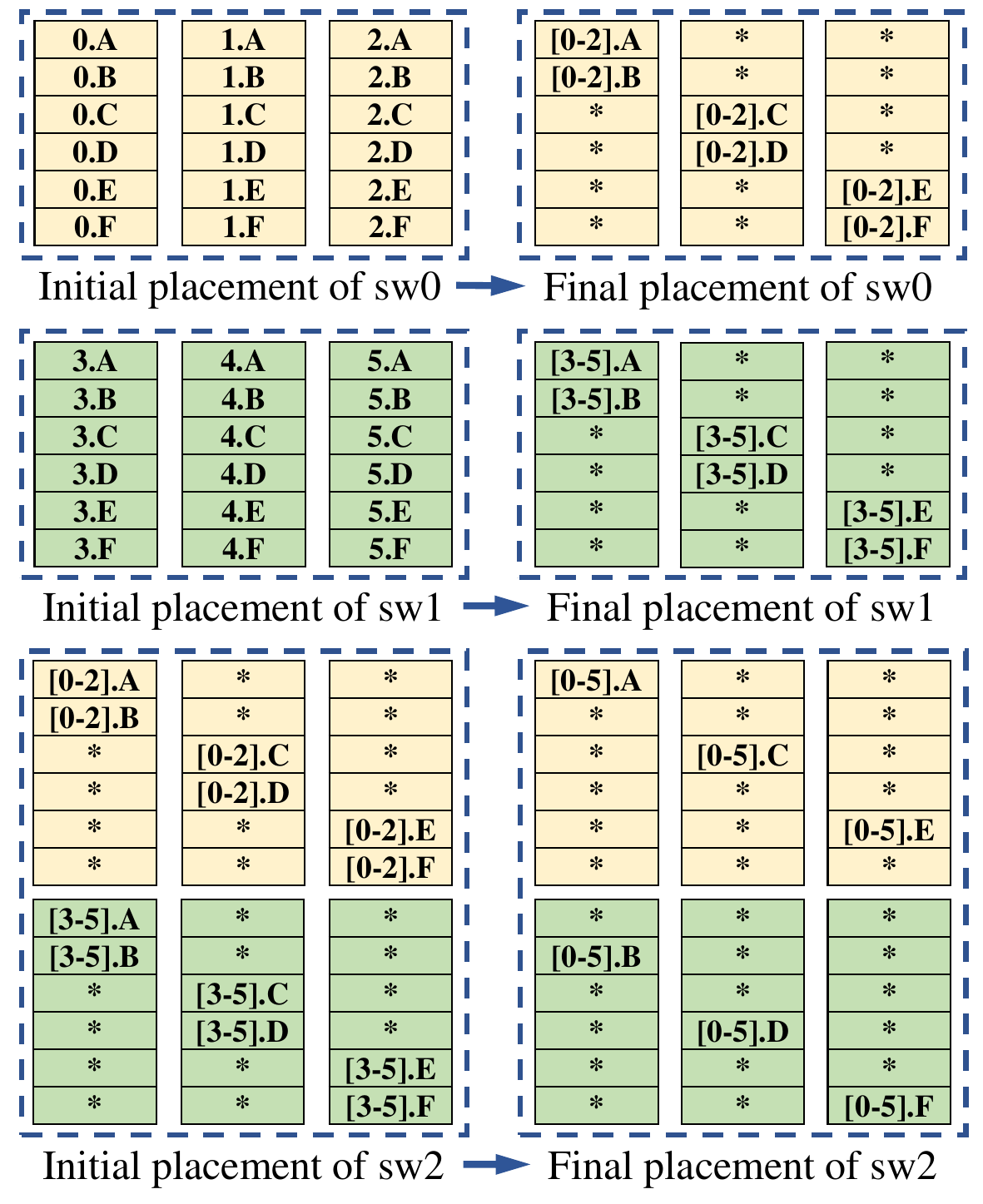}
        \vskip-8pt
        \caption{\algo Hierarchical \allr on a symmetric tree topology that shows in Figure \ref{fig:phytopo6} ($3\times2$).}
        \label{fig:hcps6}
    \end{subfigure}
    \begin{subfigure}{\linewidth}
        \centering
        \includegraphics[width=0.85\linewidth*\real{1}, height=(\linewidth*\real{0.997826086956522}*\real{1})]{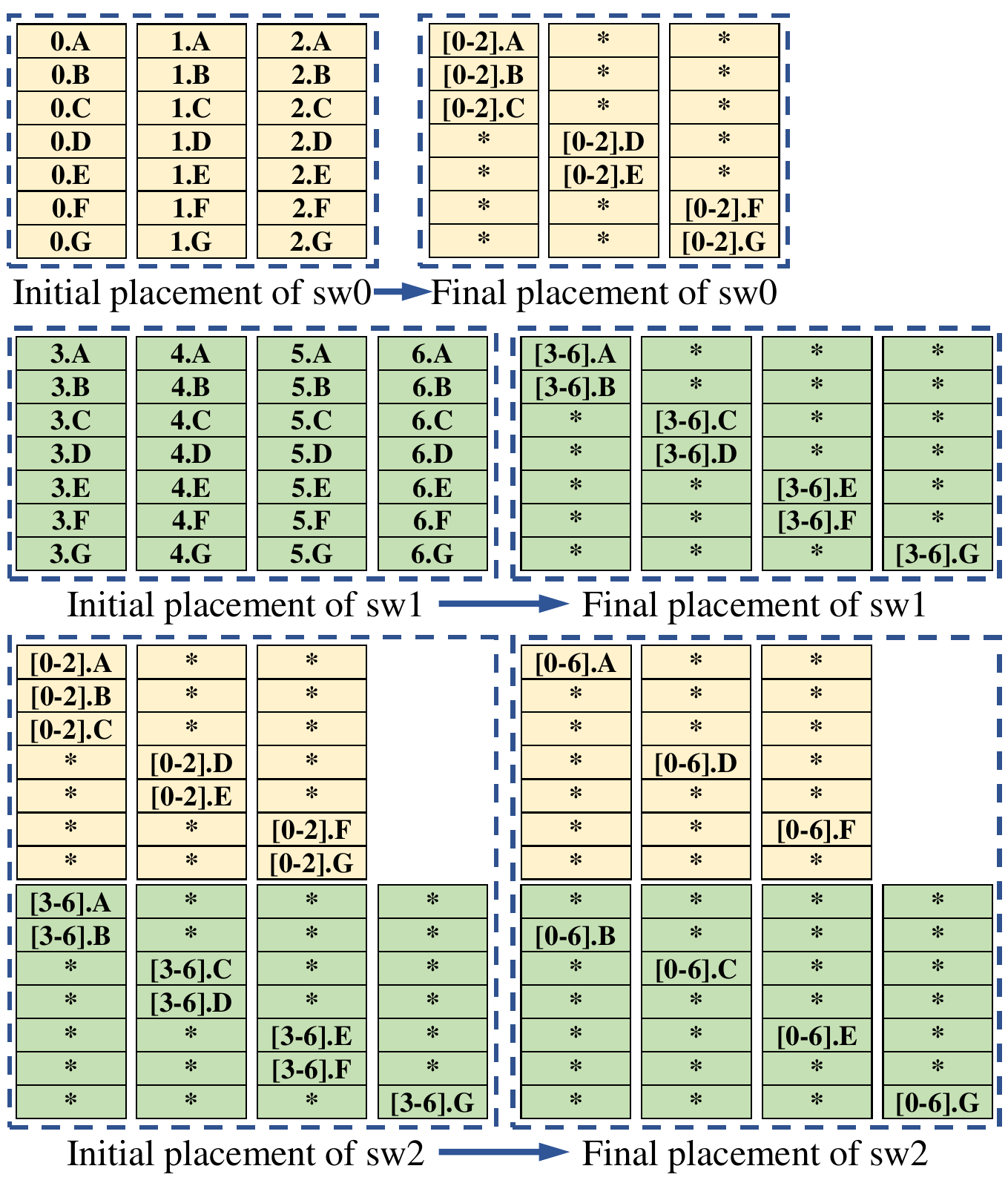}
        \vskip-8pt
        \caption{\algo Hierarchical \allr on an asymmetric tree topology that shows in Figure \ref{fig:phytopo7}. }
        \label{fig:hcps7}
    \end{subfigure}
    \caption{Examples of \algo Hierarchical \allr. Numbers are node labels and letters are data blocks. ``Placement of swX'' refers to the placements in the basic \sr plan for the swX-local sub-tree.}
    \label{fig:hcps}
\end{figure}

For example, Figure \ref{fig:phytopo} shows two different physical topologies and Figure \ref{fig:hcps} shows the basic \sr sub-plans on them. Non-asterisk cells represent the data blocks that the servers hold.


\bdpara{Final sub-plan optimizations.} With a basic sub-plan generated for each switch-local sub-tree, \algo then uses \model to optimize the basic sub-plans to produce the final plan. For each switch-local sub-tree, \algo considers two optimizations as follows. Take switch $A$ and its children $C_i$s as an example:

\begin{icompact}
    \item \textit{Data rearrangement:} This optimization aims to reduce the number of connections to prevent congestion or incast. We achieve this by aggregating the scattered results of $C_i$ to a subset of $C_i$'s children performing the switch-local \sr on $A$.
The size of the subset depends on the convergence ratio, \IE, the total bandwidth of $A$ to its children divided by that of $C_i$, to ensure that the bottleneck link can be fully utilized. To decide whether to apply this optimization, \algo leverages \model to calculate two costs for each of $C_0, C_1, \dots, C_{c-1}$: (1) without any modifications, all servers under $C_i$ transfer the scattered results out of $C_i$; (2) use \cps to rearrange the scattered results of $C_i$ to the subset, and then the servers of the subset transfer the scattered results out of $C_i$. If the latter is faster, \algo will apply this optimization. \par 

\item \textit{Plan type selection:} \algo decides which algorithm to use for the switch-local \sr operation on $A$. 
If $n_0,n_1,\dots,n_{c-1}$ are equal, this operation can adopt any of the state-of-the-art \sr algorithms listed in Table \ref{tab:genmodel_on_exist}, as their initial and final states are matched. If not, servers can only send blocks directly to the destination, which we call \emph{Asymmetric \cps}\footnote{In a standard \cps, every two nodes exchange the same amount of data, which is \emph{symmetric}. However, in this scenario, the amount of data that every two nodes exchange is not fully equal, as Figure \ref{fig:hcps7} shows, so we call it \emph{asymmetric}.}
\end{icompact}



\begin{algorithm}[tb]
    \caption{\texttt{generate\_basic\_plan($\cdots$)}}
    \label{alg:basic_plan}
    \footnotesize
    \SetKwData{node}{node}
    \SetKwData{nts}{num\_total\_servers}
    \KwIn{A node in the physical topology \node; The total number of the servers \nts}
    \BlankLine

    \If{\node.is\_server}{
        \node.basic\_plan.final\_place = \{\node: range(0,\nts)\}\;
        return\;
    }
    \For{i in \node.children}{
            generate\_basic\_plan(i)\;
    }
    taken = [false] * \nts\;
    num\_blocks = floor(\nts / num\_servers(\node))\;
    remain = num\_total\_blocks \% num\_servers(\node)\;
    \node.basic\_plan.final\_place = \{\}\;
    \For{i in \node.children}{
        \node.basic\_plan.initial\_place += i.basic\_plan.final\_place\;
        \For{server,blocks in i.basic\_plan.final\_place}{
            num\_blocks\_this\_server = num\_blocks\;
            \If{remain > 0}{
                num\_blocks\_this\_server += 1\;
                remain -= 1\;
            }
            \For{block in blocks}{
                \If{taken[block] is false}{
                    taken[block] = true\;
                    \node.basic\_plan.final\_place[server].append(block)\;
                    num\_blocks\_this\_server -= 1\;
                    \If{num\_blocks\_this\_server == 0}{
                        break\;
                    }
                }
            }
        }
    }
\end{algorithm}
\begin{algorithm}[tb]
    \caption{\texttt{generate\_final\_plan($\cdots$)}}
    \label{alg:final_plan}
    
    \footnotesize
    \SetKwData{node}{node}
    \SetKwData{nts}{num\_total\_servers}
    \KwIn{A node in the physical topology \node; Data size $S$}
    \BlankLine
    
    \If{\node.is\_server}{
        return\;
    }
    \For{i in \node.children}{
            generate\_final\_plan(i)\;
    }

    init\_place = \node.basic\_plan.init\_place\;
    final\_place = \node.basic\_plan.final\_place\;
    start\_time = 0\;

    \For{i in \node.children}{
        rearrange\_place = \textit{data placement after rearrangement}\;
        time\_origin = all\_tranfer\_out(i, i.basic\_plan.final\_place, \node, $S$)\;
        time\_rearrange = GenModel("Co-located PS", i.basic\_plan.final\_place, rearrange\_place, $S$) + all\_tranfer\_out(i, rearrange\_place, \node, $S$)\;
        \If{time\_rearrange < time\_origin}{
            i.basic\_plan.rearrange\_place = rearrange\_place\;
            i.finish\_time += GenModel("Co-located PS", i.basic\_plan.final\_place, rearrange\_place, $S$)\;
            \textit{modity} init\_place \textit{accordingly}\;
        }
        start\_time = max(start\_time, i.finish\_time)\;
    }

    best\_algo = None\;
    best\_time = INF\;
    all\_possible\_algo = []\;
    \eIf{children of \node have the same number of servers}{
        possible\_algo = \textit{all state-of-the-art algorithms}\;
    }{
        possible\_algo = ["\cps{}"]\;
    }
    \For{i in possible\_algo}{
        time = GenModel(i, init\_place, final\_place, $S$)\;
        \If{time < best\_time}{
            best\_time = time\;
            best\_algo = i\;
        }
    }

    \node.plan = best\_algo\;
    \node.finish\_time = start\_time + best\_time\;
\end{algorithm}

The corresponding pseudo code for the above \algo algorithm is shown in Algorithm \ref{alg:basic_plan} and \ref{alg:final_plan}.

We then discuss the features and advantages of our design.




\subsection{Remarks on \algo}
Compared to existing types of \allr plans (Table~\ref{tab:abcmodel_on_exist}), we believe \algo is advantageous in the following aspects.

\bdpara{Switch-local sub-tree operations are optimized independently.} On each switch, \sr only involves a switch-local sub-tree. For example, in the first step of Figure \ref{fig:hcps6}, two \sr-s are inside two switches respectively. This generates simple and symmetric traffic patterns, alleviates bottleneck links, and can leverage all the link bandwidth.

\bdpara{Using \hcps to achieve a trade-off between $\varepsilon$-optimality and $\delta$-optimality.} \model suggests a trade-off between memory access and incast overhead, and basic algorithms cannot well handle this trade-off (discussed before in Section \ref{sssec:impossible}). As a comparison, \hcps's fan-in degrees are moderate (\EG, if $N=32$, the fan-in degrees can be $8$ and $4$), thus avoiding incast and controlling memory access overhead. This is very helpful when $N$ is large.

\bdpara{Reduced congestion.} \algo attempt to reduce the number of connections to avoid congestion by data rearrangement. This is very useful when bandwidth is constrained at the upper layers of the tree (\EG, the top layer link crosses the WAN). For example, if there are two cooperating data centers, the top layer link may cross the WAN and has low bandwidth and high latency. \algo limits the number of communications for \sr of each switch-local sub-tree thus controlling the number of the flows.

\section{Implementation and Evaluation} \label{sec:impl}
We evaluate \model and \algo in this section. We seek to answer the following questions: 1) How accurate is \model? 2) What necessitates the addition of $\varepsilon$ and $\delta$? 3) Can \model help \algo achieve higher performance in both real testbed experiments and large-scale simulations? 4) How do \model and \algo perform in large-scale simulations? We summarize our findings as follows:
\bdpara{Summary of results:}
\begin{icompact}
    \item We show  that \model can correctly predict the best algorithm while the \abc model cannot. In our test scenarios, \model's maximum error is \SI{2.6}{\percent}, whereas the \abc model's is \SI{19.8}{\percent}.
    \item We implement and deploy \algo on real testbeds to verify its performance benefits. As a result, \algo outperforms the state-of-the-art algorithms. In the CPU testbed, the maximum speedup is $2.4\times$, and $1.2\times$ if RHD is excluded. In the GPU testbed, the maximum speedup over NCCL is $1.65\times$.
    \item We perform large-scale simulations to verify the accuracy and performance of \model and \algo. We find that \allr plans generated by \algo have significant advantages over state-of-the-art algorithms in all scenarios. Under different networks, the max speedup is between $4.9\times$ and $7.4\times$. To assist reproduction of our results, we release the code for the simulation\footnote{\githuburl}.
\end{icompact}

\subsection{\model Accuracy}
\label{ssec:evalcm} 
In this section, we evaluate \model on several state-of-the-art \allr algorithms. We implement \ring (which NCCL usually uses), RHD (which MPI usually uses), \cps, and \hcps with Open MPI v4.1.1\@\cite{openmpi}. 

We use the same testbed setting as in Section~\ref{sec:genmodel}. On our testbed, \model is parameterized by feeding \cps benchmarks ranging from $N=2$ to $15$. We first test the accuracy of \model, then take a deeper look to analyze the impact of the five terms in \model.  \par 

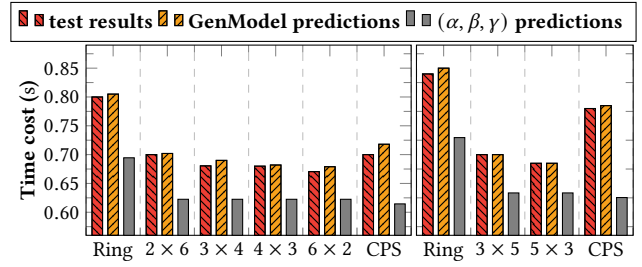
\begin{figure}[tb]
\hypersetup{linkcolor=black}
\footnotesize
\centering
\ref{legend:model_prediction}
\begin{tikzpicture}
\begin{groupplot}[
        group style={
            group size=2 by 1,
            horizontal sep=2pt,
            x descriptions at=edge bottom,
            ylabels at=edge left,
        },
        scale only axis,
        height=72pt,
        ymin=0.56,
        ymax=0.89,
        xtick distance=1,
        ytick distance=0.05,
        minor x tick num=1,
        minor y tick num=1,
        xminorgrids,
        grid style=dashed,
        ylabel shift=-3pt,
        ybar,
        y tick label style={
            /pgf/number format/.cd,
                fixed,
                fixed zerofill,
                precision=2,
            /tikz/.cd
        },
        xlabel style={at={(0.5,-0.2)}},
        xlabel={\phantom{x}},
        xtick align=inside,
    ]
    \nextgroupplot[
        ylabel=Time cost (\si{\s}),
        legend to name=legend:model_prediction,
        legend columns=-1,
        width=0.51\linewidth,
        bar width=4pt,
        symbolic x coords={Ring,$2\times6$, $3\times4$, $4\times3$, $6\times2$, CPS},
    ]
    \addplot+[ybar, ACMRed , draw=black, postaction={pattern={north west lines}, pattern color=black}]
    coordinates
    {
        (CPS,0.7) ($6\times2$,0.6705) ($4\times3$,0.68) ($3\times4$,0.6805) ($2\times6$,0.7) (Ring,0.8)
    };
    \addplot+[ybar, ACMOrange , draw=black, postaction={pattern={north east lines}, pattern color=black}]
    coordinates
    {
        (CPS,0.718) ($6\times2$,0.679) ($4\times3$,0.682) ($3\times4$,0.69) ($2\times6$,0.702) (Ring,0.805)
    };
    \addplot+[ybar, gray, draw=black]
    coordinates
    {
        (CPS,0.61450333) ($6\times2$, 0.62250333) ($4\times3$,0.62250333) ($3\times4$,0.62250333) ($2\times6$,0.62250333) (Ring, 0.69450333)
    };
    \addlegendentry{test results}
    \addlegendentry{\model predictions}
    \addlegendentry{$(\alpha,\beta,\gamma)$ predictions}
    \nextgroupplot[
        width=0.34\linewidth,
        bar width=4pt,
        symbolic x coords={Ring,$3\times5$,$5\times3$,CPS},
        ymajorticks=false,
        enlarge x limits=0.1675,
    ]
    \addplot+[ybar, ACMRed , draw=black, postaction={pattern={north west lines}, pattern color=black}]
    coordinates
    {
        (CPS,0.78) ($5\times3$,0.685) ($3\times5$,0.7) (Ring,0.84)
    };
    \addplot+[ybar, ACMOrange , draw=black, postaction={pattern={north east lines}, pattern color=black}]
    coordinates
    {
        (CPS,0.785) ($5\times3$,0.685) ($3\times5$,0.7) (Ring,0.85)
    };
    \addplot+[ybar, gray, draw=black]
    coordinates
    {
        (CPS,0.625530666666667) ($5\times3$,0.633530666666667) ($3\times5$,0.633530666666667) (Ring, 0.729530666666667)
    };

\end{groupplot}
\end{tikzpicture}
\vskip-8pt
\caption{\model predictions of algorithm time cost on 12 nodes (left) and 15 nodes (right). ``Ring'' is for \ring and ``CPS'' for \cps.}
\label{fig:costmodel}
\end{figure}



We test \ring, \cps, and \hcps under the scale of $12$ and $15$ processors. \hcps is denoted by $a\times b$ meaning that the number of steps $m=2$ and fan-in degrees $f_0=a,f_1=b$. Figure~\ref{fig:costmodel} shows the actual cost, \model predicted cost, and the \abc model predicted cost. The results show that the error of \model is within \SI{2.6}{\percent}, demonstrating the prediction ability of \model. As a comparison, the ($\alpha,\beta,\gamma$) model can neither estimate the cost nor select the optimal algorithm, and its error is up to \SI{19.8}{\percent}. Furthermore, the results also confirm our prediction in Section \ref{ssec:fmltmodel} that using hierarchical \allr in one layer may be beneficial. \par 

\begin{figure}[tb]
    \hypersetup{linkcolor=black}
    \centering
    \ref{breakdown_legend}
    \begin{tikzpicture}
        \begin{groupplot}[
            group style={
                group size=2 by 1,
                horizontal sep=22pt,
                x descriptions at=edge bottom,
            },
            symbolic x coords={Ring, $2\times6$,$3\times4$,$4\times3$,$6\times2$,CPS},
            xtick distance=1,
            ybar,
            ymajorgrids,
            width=0.57\linewidth,
            height=0.50\linewidth,
            ymin=0,
             y tick label style={
                /pgf/number format/.cd,
                    fixed,
                    fixed zerofill,
                    precision=2,
                /tikz/.cd
            },
            ticklabel style={font=\scriptsize},
            xtick align=inside,
            ylabel shift=-4pt,
            minor x tick num=1,
            xminorgrids,
            x grid style=dashed,
        ]
        \nextgroupplot[
            ylabel = {Time cost (\si{\s})},
            ytick distance=0.1,
            legend entries={Communication, Calculation},
            legend columns=-1,
            legend to name={breakdown_legend},
            bar width=5pt,
        ]
        \addplot+[ybar, black, pattern={crosshatch}, pattern color=ACMDarkBlue] coordinates {
            (Ring, 0.660933) ($2\times6$,0.5754909) ($3\times4$,0.5760999) ($4\times3$,0.5713501) ($6\times2$,0.5769914) (CPS,0.6261722)
        };
        \addplot+[ybar, draw=black, fill=ACMRed] coordinates {
            (Ring,0.087144) ($2\times6$,0.0645091) ($3\times4$,0.0539001) ($4\times3$,0.0486499) ($6\times2$,0.0430086) (CPS,0.0338278)
        };
        \nextgroupplot[
            ytick distance=0.05,
            ymax=0.19,
            bar width=5pt,
        ]
        \addplot+[ybar, black, pattern={crosshatch}, pattern color=ACMDarkBlue] coordinates {
            (Ring, 0.130613) ($2\times6$,0.1181375) ($3\times4$,0.117556) ($4\times3$,0.118159) ($6\times2$,0.119825833) (CPS,0.128188)
        };
        \addplot+[ybar, draw=black, fill=ACMRed] coordinates {
            (Ring, 0.087) ($2\times6$,0.064774) ($3\times4$,0.058483217) ($4\times3$,0.0500399) ($6\times2$,0.043595283) (CPS,0.0340864)
        };
        \end{groupplot}
    \end{tikzpicture}
    \caption{Time cost break-down of different algorithms on 12 processors in \SI{10}{Gbps} (left) and \SI{100}{Gbps} (right) network. ``Ring'' is for \ring and ``CPS'' for \cps. }
    \label{fig:breakdown}
\end{figure}
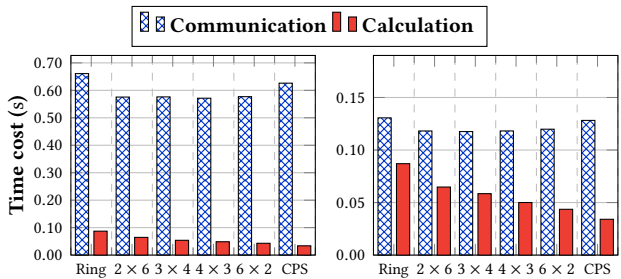
\begin{figure}[t]
    \hypersetup{linkcolor=black}
    \centering
    \phantom{WWW}\ref{legend:model_five}\\
    \begin{tikzpicture}
        \begin{axis}[
            xtick distance=1,
            ytick distance=10,
            minor y tick num=1,
            minor x tick num=1,
            xminorgrids,
            ylabel={Proportion (\si{\percent})},
            ybar stacked,
            bar width=9pt,
            ymajorgrids,
            height=0.5\linewidth,
            width=0.8\linewidth,
            ymin=70, ymax=100,
            grid style=dashed,
            xlabel shift=-7pt,
            xtick pos=bottom,
            symbolic x coords={Ring, $2\times6$, $3\times4$, $4\times3$,$6\times2$,CPS},
            legend entries={bandwidth\&computation, bandwidth contention, memory I/O, latency},
            legend columns=2,
            legend to name={legend:model_five},
            reverse legend,
        ]  
        \addplot+[ybar, ACMDarkBlue, draw=black] coordinates {
            (Ring,75.66788655) ($6\times2$,89.50229227) ($4\times3$,88.65366063) ($3\times4$,87.82097073) ($2\times6$,86.20165249) (CPS,84.82010498) 
        }; 
        \addplot+[ybar, ACMGreen, draw=black, postaction={pattern=north east lines, pattern color=black}] coordinates { 
            (Ring,0) ($2\times6$,0) ($3\times4$,0) ($4\times3$,0) ($6\times2$,0) (CPS,8.164502084)
        }; 
        \addplot+[ybar, ACMRed , draw=black, postaction={pattern={north west lines}, pattern color=black}] coordinates {
            (Ring,13.35315645) ($6\times2$,8.136572024) ($4\times3$,9.007591187) ($3\times4$,9.86224805) ($2\times6$,11.52428509) (CPS,5.896584838)
        }; 
        \addplot+[ybar, ACMYellow, draw=black] coordinates {
            (Ring,10.978957) ($6\times2$,2.361135706) ($4\times3$,2.338748185) ($3\times4$,2.316781219) ($2\times6$,2.274062423) (CPS,1.118808096)
        }; 
        \end{axis}
    \end{tikzpicture}
    \caption{Time cost break-down by \model for different algorithms on 12 processors in \SI{10}{Gbps} network. ``Ring'' is for \ring and ``CPS'' for \cps.}
    \label{fig:model_five}
\end{figure}
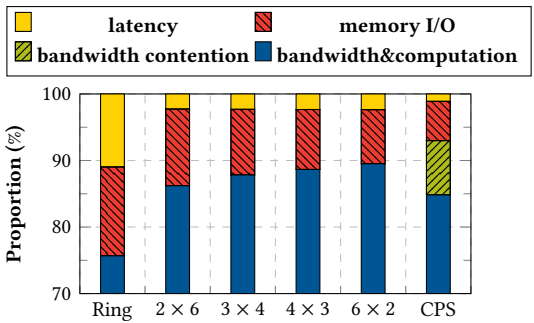

Then we break the time cost down. We set up a dedicated timer for the reduce function which involves the $\gamma$ and the $\delta$ terms (called \emph{calculation}). The time cost of the other three terms (called \emph{communication}) is obtained by subtracting the calculation time from the total time. We set the link speed to \SI{100}{Gbps} to better show the impact of memory access costs. Figure~\ref{fig:breakdown} shows the results. First, with the increase of the first step's fan-in degree, the calculation cost decreases monotonically. In the \SI{100}{Gbps} network, compared to \ring, \cps reduces the calculation cost by \SI{61}{\percent}. This confirms the existence of the memory access overhead and algorithms can benefit from reducing it. Second, the communication costs of \ring and \cps are higher than others: \ring has a high latency ($\alpha$) term, and \cps has a high incast term ($\varepsilon$). \ring has too many communication steps so the latency is high; \cps has many-to-one communications so the incast term ($\varepsilon$) emerges. With \model, all these phenomena can be reasonably explained.

In more detail, Figure~\ref{fig:model_five} leverages our model to analyze the impact of each term, which agrees with the break-down tests in Figure~\ref{fig:breakdown}. As the sums of bandwidth and computation cost (\IE, $\beta\text{-term} + \gamma\text{-term}$) of different algorithms are the same in theory, the larger its proportion, the lower the total overhead. The other three components show a clear trade-off: with the increase of the fan-in degree, the memory access and latency terms decrease while incast overhead increases, which generates an optimal choice of $6\times2$. These results are consistent with our previous analysis, confirming the necessity of adding the new two terms to the existing model. \par 





\subsection{Testbed Experiments for \algo}\label{ssec:testbedexpr}
\begin{table}[t]
    \centering
    \small
    \caption{Test results for GenTree on CPU testbed. \textit{\algo} means plans generated by \algo; same below.}
    \label{tab:gentreereal}
    \begin{tabular}{c|*{3}{c}}
        \hline
        & \multicolumn{3}{c}{Time cost (s) with \#servers} \\ \cline{2-4}
        Algorithm & \num{8} & \num{12} & \num{15} \\ \hline 
        \algo & $\textbf{0.647}$ & $\textbf{0.620}$ & $\textbf{0.632}$ \\
        \cps & $\textbf{0.647}$ & $0.660$ & $0.731$ \\
        \ring & $0.719$ & $0.748$ & $0.758$ \\
        RHD & $0.736$ & $1.520$ & $1.521$ \\ \hline
    \end{tabular}
\end{table}

We use \algo to generate \allr plans on our real testbeds, and then compare these plans to the state-of-the-art \allr plans. To ensure the universality of the test results, we establish two testbeds. The CPU testbed shares the same setting as in Section~\ref{sec:genmodel}, which contains 15 servers connected to one single switch and the reduce operation is done by CPU. Each server has only one NIC and it is set to the speed of \SI{10}{Gbps}. The GPU testbed has 8 DGX-A100 servers randomly chosen from a pod of Fat-Tree network~\cite{fattree}. Each server has 8 NVIDIA A100 GPUs and 4 NICs on \SI{200}{Gbps}. The convergence ratio of edge switches is $1\mathbin{:}1$. The reduce operation is done by GPU. RDMA or GDR is enabled in all scenarios. All tests are repeated 100 times to avoid network noise.

In the CPU testbed, we implement and perform \allr with Open MPI. The data size is set to \num{1e8}, when $8$ servers are connected, \algo chooses \cps; for $12$ servers, \algo chooses $6\times2$ \hcps; for $15$ servers, \algo chooses $5\times3$ \hcps. This is because \model suggests that when the fan-in degree is greater than $w_t$ (in this network, $w_t=9$), the incast overhead will emerge. \par 

Results of respective plans are shown in Table \ref{tab:gentreereal}. \algo-generated plans successfully outperforms other plans with a maximum speedup $2.4\times$ ($1.2\times$ excluding RHD). We can infer that (1) compared to RHD and \ring, \algo is advantageous because it reduces memory access overhead, and (2) compared to \cps, \algo avoids incast; (3) RHD is suited for networks with a power-of-two number of servers, and if this condition is not met, overheads increase significantly. 

In the GPU testbed, we implement and perform \allr with NCCL~\cite{jeaugey2017nccl} and ps-lite~\cite{pslite}. NCCL is used to provide GPU compatibility and high-performance intra-machine communication; ps-lite is used to support customized inter-machine communication. For $n$ servers, \algo chooses $8\times n$ hierarchical \allr plan, in which the first step is \texttt{ncclAllReduce} (intra-machine) and the second step is \cps (inter-machine). This is because \model takes into account the physical topology and finds that there is no need for additional layering in inter-machine communication: the fan-in degree there is less than the threshold $w_t$, which is different from the CPU cluster. We take NCCL~\cite{jeaugey2017nccl} as the baseline because it is the most widely-used communication library on NVIDIA GPUs. Others are not compared, such as (1) SCCL/TACCL\cite{SCCL,taccl}, which fails to synthesize algorithms in \SI{72}{h} on our testbed; (2) $P^2$\cite{2022xie_synthesizing}, which focuses on hybrid parallel strategies of machine learning and is not comparable to pure \allr primitive.

\begin{table}[t]
    \centering
    \caption{Test results for GenTree on GPU testbed}
    \label{tab:gentreerealgpu}
    \small
    \begin{tabular}{c|c|*{4}{c}}
        \hline
        & & \multicolumn{4}{c}{Time cost (ms) with data size (float)} \\ \cline{3-6}
        \#GPUs & Algorithm & \num{1e7} & \num{3.2e7} & \num{1e8} & \num{3.2e8} \\ \hline 
        \multirow{2}*{16} & \algo & $\textbf{0.764}$ & $\textbf{1.677}$ & $\textbf{5.058}$ & $\textbf{15.501}$ \\
        & NCCL & ${0.941}$ & $2.695$ & $8.170$ & $25.606$ \\ \hline
        \multirow{2}*{32} & \algo & $\textbf{0.842}$ & $\textbf{2.340}$ & $\textbf{7.030}$ & $\textbf{22.343}$ \\
        & NCCL & ${1.011}$ & $3.163$ & $9.081$ & $27.978$ \\ \hline
        \multirow{2}*{64} & \algo & $\textbf{0.971}$ & $\textbf{2.668}$ & $\textbf{8.093}$ & $\textbf{25.716}$ \\
        & NCCL & ${1.149}$ & $3.243$ & $9.886$ & $31.049$ \\ \hline
    \end{tabular}
\end{table}


Results are shown in Table \ref{tab:gentreerealgpu}. \algo plans show promising performance with a maximum speedup $1.65\times$ over NCCL. This advantage weakens ($1.65\times \to 1.22\times$) when the number of servers increases ($2\to8$), which is because of the growth of inter-machine communications data traffic $1/2 \to 7/8$). As the number of servers increases, the speedup ratio will converge to approximately $1.2\times$.



\subsection{Large-scale Simulations for \algo}\label{ssec:evalsim}

Limited by the size of our testbed, we have to rely on simulations for large-scale experiments.
In this section, we evaluate \algo on simulators, considering both the computation and the communication overheads. Computation time is derived from the $\gamma\text{-term}$ and the $\delta\text{-term}$ in \model. Communication time is obtained from a custom-made flow-level network simulator which is aware of the incast problem. This is because (1) packet-level network simulators such as ns3~\cite{ns3} consume too much time on large networks; (2) we do not need the level of details provided by the packet-level simulator. We release the source code for reproduction of our results.
The parameters of our simulator are obtained by the fitting methodology described in Section~\ref{ssec:fitting}, as shown in Table~\ref{tab:gtcoe}.\par

\begin{table}[t]
    \centering
    \caption{Parameters in GenModel for different physical topology.}
    \label{tab:gtcoe}
    \begin{adjustbox}{max width=\linewidth}
    \begin{tabular}{c | c| c| c| c| c| c}
        \hline
        \makecell{Type} & $\alpha$ & $\beta$ & $\gamma$ & $\delta$ & $\varepsilon$ & $w_t$ \\ \hline
        Cross DC & \num{3.00e-2} & \num{6.40e-9} & / & / & \num{6.00e-11} & \num{9e0}\\ \hline
        Root SW & \num{6.58e-3} & \num{6.40e-10} & / & / & \num{6.00e-12} & \num{9e0}\\ \hline
        Middle SW & \num{6.58e-3} & \num{6.40e-9} & / & / & \num{1.22e-10} & \num{9e0}\\ \hline
        Server & \num{6.58e-3} & / & \num{6.00e-10} & \num{1.87e-10} & / & / \\ \hline
    \end{tabular}
    \end{adjustbox}
\end{table}

\begin{figure}[tb]
    \centering
    \includegraphics[width=\linewidth]{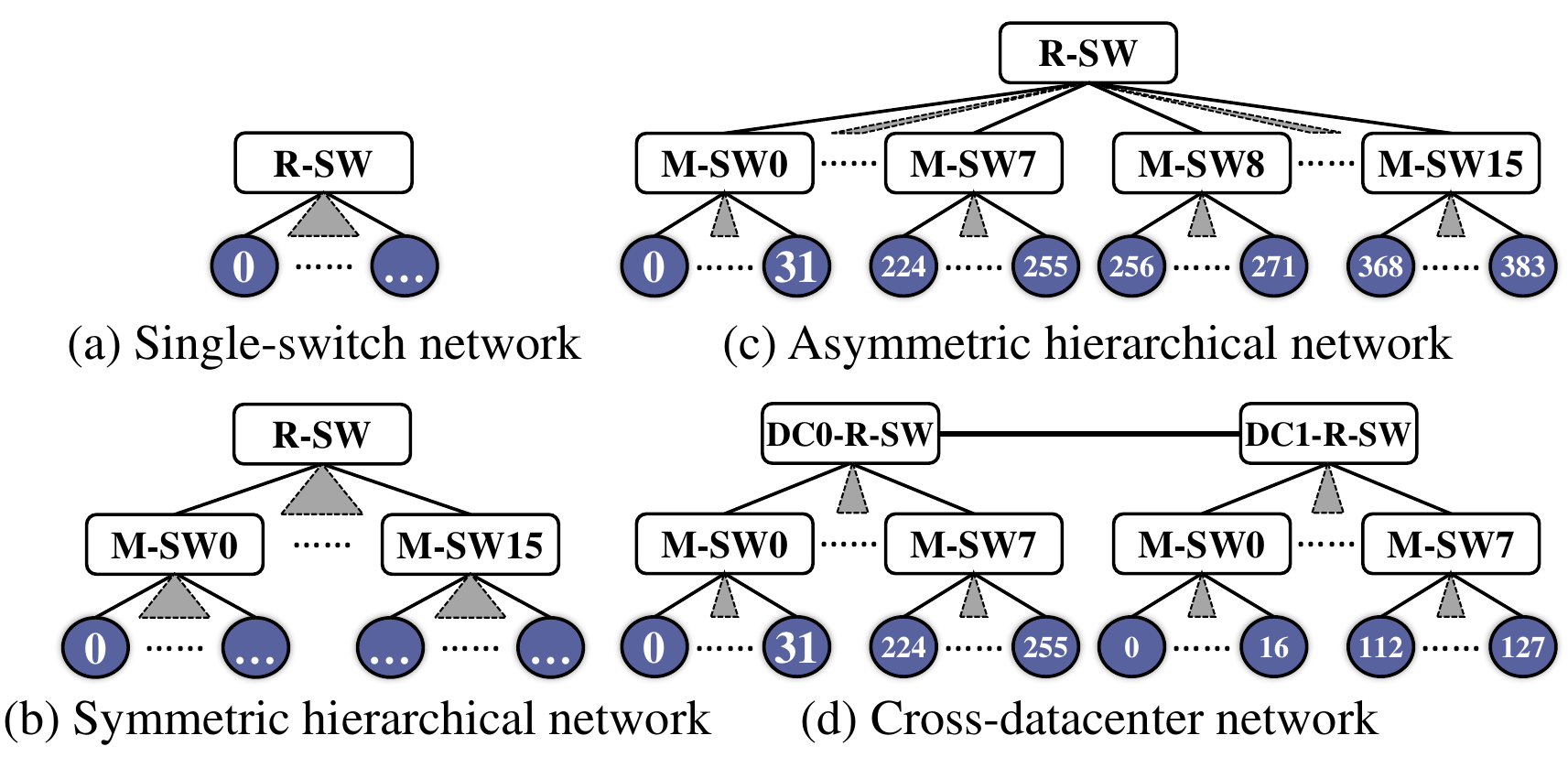}
    \caption{Four Representative physical topologies that are used to evaluate \algo. ``R-SW'' is for ``Root Switch'' and ``M-SW'' is for ``Middle-layer Switch''.}
    \label{fig:fourphytopo}
\end{figure}

\bdpara{Physical topologies:}
As listed below, we set up four representative physical topologies to evaluate \algo. They are shown in Figure~\ref{fig:fourphytopo}.
\begin{icompact}
    \item[1.] Single-switch network. This is a common topology for an in-rack cluster and an essential component for building large-scale networks. We use two instances: (SS24) $24$ servers connected to a switch; and (SS32) $32$ servers connected to a switch.
    \item[2.] Symmetric hierarchical network. This is a common single-root tree topology with $16$ middle-layer switches connected to the root switch. We vary the number of servers connected to middle-layer switches from $24$ ($384$ servers in total, SYM384) to $32$ ($512$ servers in total, SYM512).
    \item[3.] Asymmetric hierarchical network. We use the instance that there are $16$ middle-layer switches connected to the root switch, and configure half of the middle-layer switch to connect to $32$ servers, and the other half $16$. There are $32\times 8 + 16\times 8=384$ servers in total (ASY384). 
    \item[4.] Cross-datacenter network. This topology features a top-layer link that has low bandwidth and high latency. We use the instance that, (1) in one data center, there are $8$ middle-layer switches connected to the root switch and $32$ servers connected to each of the middle-layer switches; (2) in the other data center, there are $8$ middle-layer switches connected to the root switch and $16$ servers connected to each of the middle switches; (3) the two data centers' root switches are connected through one link. There are $32\times8 + 16\times 8=384$ servers in total (CDC384).  
\end{icompact} \par 

\bdpara{\algo generated \allr plans:}
Using the above topologies and different data sizes (\num{1e7}, \num{3.2e7}, and \num{1e8}) as input, \algo generates various \allr plans for different switches, shown in Table~\ref{tab:gtplan}.

\begin{table}[tb]
    \centering
    \small
    \caption{\allr plans selected by \algo. CPS is for \cps, $m\times n$ for $m\times n$ \hcps, ACPS for Asymmetric \cps.}
    \label{tab:gtplan}
    
    \begin{adjustbox}{max width=\linewidth}
    \begin{tabular}{c|c|*{3}{c}}
        \hline
        &  \multirowcell{2}{Switch-local\\sub-tree} & \multicolumn{3}{c}{Plan on data size (float)} \\ \cline{3-5}
        \makecell{Network} & & \num{1e7} & \num{3.2e7} & \num{1e8} \\ \hline
        SS24 & Root SW & CPS & $8\times3$ & $8\times3$ \\ \hline
        SS32 & Root SW & $8\times4$ & $8\times4$ & $8\times4$ \\ \hline
        \multirow{2}*{SYM384} & Middle SW & CPS & $8\times3$ & $8\times3$ \\ 
        & Root SW &CPS & $8\times2$ & $8\times2$ \\ \hline
        \multirow{2}*{SYM512} & Middle SW & CPS & $8\times4$ & $8\times4$ \\ 
        & Root SW & CPS & $8\times2$ & $8\times2$ \\ \hline
        \multirow{3}*{ASY384} & Middle SW 0-7 & CPS & $8\times4$ & $8\times4$ \\ 
        & Middle SW 8-15 & CPS & $8\times2$ & $8\times2$ \\
        & Root SW & ACPS & ACPS & ACPS \\ \hline
        \multirow{5}*{CDC384} & DC0 Middle SW & CPS & $8\times4$ & $8\times4$ \\ 
        & DC0 Root SW & CPS & CPS & CPS \\
        & DC1 Middle SW & CPS & $8\times2$ & $8\times2$ \\
        & DC1 Root SW & CPS & CPS & CPS \\ 
        & Cross DC & ACPS & ACPS & ACPS \\\hline
    \end{tabular}
    \end{adjustbox}
\end{table}

\bdpara{Baseline \allr plans:}
Baseline \allr plans are \ring (which NCCL usually uses), RHD (which MPI usually uses) and \cps (which PS-based \allr usually uses). Since RHD is not suitable for non-power-of-two networks, in the following, we only evaluate RHD when the number of servers is power-of-two.

\begin{table}[t]
    \centering
    \caption{Large-scale simulation results for GenTree. \algo{*} is the special plan without data rearrangement.}
    \label{tab:gentreesimu}
    \small
    \begin{tabular}{c|c|*{3}{p{3.8em}}}
        \hline
        & & \multicolumn{3}{c}{Time cost (s) on data size (float)} \\ \cline{3-5}
        Topo & Algorithm & \num{1e7} & \num{3.2e7} & \num{1e8} \\ \hline

        \multirow{3}*{SS24} & \algo & $\textbf{0.203}$ & $\textbf{0.503}$ & $\textbf{1.404}$ \\
        & Ring-Allr & $1.082$ & $1.376$ & $2.288$ \\
        & C-PS & $\textbf{0.203}$ & $0.562$ & $1.673$ \\ \hline
        \multirow{4}*{SS32} & \algo & $\textbf{0.213}$ & $\textbf{0.507}$ & $\textbf{1.417}$ \\
        & RHD & $0.337$ & $0.644$ & $1.593$ \\
        & Ring-Allr & $1.399$ & $1.697$ & $2.617$ \\
        & C-PS & $0.223$ & $0.628$ & $1.879$ \\ \hline

        \multirow{3}*{SYM384} & \algo & $\textbf{0.503}$ & $\textbf{1.287}$ & $\textbf{3.575}$ \\
        & \ring & $2.943$ & $3.627$ & $5.742$ \\
        & \cps & $2.274$ & $7.132$ & $22.148$ \\ \hline

        \multirow{4}*{SYM512} & \algo & $\textbf{0.639}$ & $\textbf{1.627}$ & $\textbf{4.638}$ \\
        & RHD & $0.896$ & $1.853$ & $4.812$ \\
        & \ring & $3.571$ & $4.479$ & $7.285$ \\
        & \cps & $3.479$ & $10.989$ & $34.200$ \\ \hline

        \multirow{3}*{ASY384} & \algo & $\textbf{0.570}$ & $\textbf{1.593}$ & $\textbf{4.670}$ \\
        & \ring & $3.043$ & $3.947$ & $6.741$ \\
        & \cps & $2.052$ & $6.421$ & $19.925$ \\ \hline

        \multirow{4}*{CDC384} & \algo & $\textbf{2.427}$ & $\textbf{8.299}$ & $\textbf{25.388}$ \\
        & \algo{*} & $4.484$ & $13.927$ & $43.116$ \\
        & \ring & $8.513$ & $17.329$ & $44.580$ \\
        & \cps & $11.890$ & $37.799$ & $117.882$ \\ \hline
    \end{tabular}

\end{table}

\bdpara{Simulation results:} Results are shown in Table \ref{tab:gentreesimu}. \algo-generated plans show significant advantages over state-of-the-art algorithms in all scenarios. Observations are:
\begin{icompact}
    \item[1.] The max speedup (1) over the three data sizes is between $5.8\times$ and $7.4\times$; (2) in different networks is between $4.9\times$ and $7.4\times$. 
    \item[2.] When data size is small, the $\alpha$ dominates. Therefore, \algo adopts \cps instead of the hierarchical version, and \ring is slow due to having more communication steps.
    \item[3.] \algo excels when the network becomes complex. ($1.4\times$ \textasciitilde $5.3\times$ in ASY384 and $1.8\times$\textasciitilde $4.9\times$ in CDC384). State-of-the-art algorithms can not adapt to complex topologies accordingly. \algo successfully adapts to asymmetric hierarchical networks, which previous solutions cannot. 
    \item[4.] Data rearrangement saves \SI{54}{\percent}\textasciitilde\SI{60}{\percent} of time in the cross-datacenter scenario. \algo successfully avoids severe congestion on the inter-datacenter link. 
\end{icompact}

\bdpara{Summary:} The accurate \model enables \algo to generate appropriate \allr plans in diverse scenarios, and the generated plans achieve equivalent or superior performance.




\section{Related Work}\label{sec:related_work} 
\bdpara{\allr Algorithms.} \ring \cite{ringbaidu, patarasuk2009bandwidth} constructs one ring and processors only communicate with their neighbors, which results in high latency and a long dependency chain. RHD \cite{thakur2005optimization} constructs complete binary trees and processors exchange data pairwise. It has moderate latency but works badly when the number of processors is non-power-of-two. \cite{kolmakov2020generalization} proposes another non-power-of-two patch to RHD, but it will break the independence of full-meshes and makes RHD lose support for hierarchical physical architecture. Recursive Multiplying \cite{RUEFENACHT201724} is a generalization of Recursive Doubling but they are both not bandwidth-optimal.

\bdpara{Topology-aware \allr.} BytePS \cite{jiang2020unified} is designed to leverage spare CPU and bandwidth resources to accelerate distributed DNN training. HiPS \cite{geng2018hips} constructs hierarchical \allr plan on specific server-centric topologies to accelerate distributed machine learning. RAT \cite{wan2020rat} constructs trees resembling physical topology, hence reducing cross-region traffic and shortening the dependency chain. However, RAT enumerates and constructs all possible trees for load balancing, making it impractical. BlueConnect~\cite{cho2019blueconnect} breaks the \allr process down to several concurrent \sr and \op{AllGather} operations, which inspires \algo. However, its algorithm is restricted to Ring \allr, which we found to be not optimal in our evaluations.
$P^2$\cite{2022xie_synthesizing} targets for hybrid parallel strategy in large-scale deep learning. It decides how to best place tensor shards onto several devices and synthesis reduction strategy according to the placement. It is possible for $P^2$ to adopt \algo to improve its performance further. Some other work~\cite{sapio2019scaling,lao2021atp} focuses on in-network aggregation which uses programmable network devices to accelerate the reduce operation. 

\par


\bdpara{Cost Model.} Cost models are simple equations, formulas or functions used to measure, quantify and estimate the time cost of \allr. The \abc model has been introduced into the communication field by Hockney \cite{1994The}, which is used to characterize the communication cost, further used by Thakur et al. \cite{thakur2005optimization} and Cai et al. \cite{SCCL} for communication optimization. SCCL~\cite{SCCL} uses SMT solver to synthesize the optimal algorithm based on the \abc model. Due to the NP-hardness, as we have tested, SCCL commonly applies to intra-machine topologies and consumes \textit{unacceptable time} to synthesize the best \allr algorithm on large clusters. Based on SCCL, TACCL~\cite{taccl} improves scalability, but is still not applicable to large clusters, also uses the \abc model and SMT solver to synthesize \allr algorithms with an integer linear programming (ILP) encoding. Compared to SCCL, TACCL has better scalability, but is still not applicable to large clusters due to the NP-hardness.

\bdpara{RDMA Congestion Control.} Some prior work tries to improve RDMA congestion control mechanism and reduce the overhead of bandwidth contention. DCQCN \cite{zhu2015congestion-dcqcn} relies on ECN and the transmission rate is regulated according to the number of ECNs. Timely \cite{mittal2015timely} is based on RTT and the transmission rate is tuned by RTT gradients when RTT is between the high and the low thresholds. HPCC \cite{li2019hpcc} requires hardware support for in-band network telemetry.

\section{Conclusion}\label{sec:conclusion}
We propose an accurate cost model \model for \allr. We identify two neglected factors (memory access and incast) on modern clusters. Using \model, we proceed to design \algo, a topology-aware algorithm that generates highly efficient \allr plans on tree-based physical topology. Experiments show that \model can well characterize the real system overheads and \algo has a considerable improvement over the existing state-of-the-art solutions. 

We release a benchmarking toolkit to fit \model to new clusters and our simulator implementation at \githuburl.



\clearpage

\bibliographystyle{ACM-Reference-Format}
\bibliography{ref}

\end{document}